\documentclass[aps,prl,twocolumn,floatfix]{revtex4-1}
\usepackage{float}
\usepackage{graphicx}
\usepackage{amssymb}
\usepackage{amsfonts}
\usepackage{amsmath}
\usepackage{relsize}
\usepackage{color}
\usepackage{xcolor}
\usepackage{comment}

\begin{document}

\title{Disorder induced transition from type-I to type-II superconductivity in the Dirac semimetal PdTe$_{2}$}
\author{M. V. Salis} \email{m.v.salis@uva.nl}
\author{J. P. Lorenz}
\author{Y. K. Huang}
\author{A. de Visser} \email{a.devisser@uva.nl}
\affiliation{Van der Waals - Zeeman Institute, University of Amsterdam, Science Park 904, 1098 XH Amsterdam, The Netherlands}

\date{\today}

\begin{abstract}
We report a doping study directed to intentionally induce disorder in PdTe$_{2}$~by the isoelectronic substitution of Pt. Two single-crystalline batches Pd$_{1-x}$Pt$_x$Te$_2$ have been prepared with nominal doping concentrations $x=0.05$ and $x=0.10$. Sample characterization by energy dispersive x-ray spectroscopy (EDX) revealed Pt did not dissolve homogeneously in the crystals. For the nominal value $x=0.10$ small single crystals cut from the batch appeared to have $x=0.09$, as well as the non stoichiometric composition Pd$_{0.97}$Pt$_{<0.004}$Te$_{2.03}$. Magnetic and heat capacity measurements demonstrate a transition from type-I to type-II superconducting behavior upon increasing disorder. From transport measurements we calculate a residual resistivity $\rho_0 = 1.4~\mu\Omega$cm suffices to turn PdTe$_{2}$~into a superconductor of the second kind.
\end{abstract}

\maketitle

\section{introduction}
Recently, interest in transition metal dichalcogenides has increased significantly due to their extraordinary electronic properties. Notably, the opportunity to realize novel quantum states arising from the topologically non-trivial band structure, as found by density functional theory~\cite{Soluyanov2015,Huang2016,Yan2017,Bahramy2018}, attracts much attention. The formation of both type-I and type-II bulk Dirac cones has been predicted~\cite{Bahramy2018}. Of special interest in this family is the semimetal PdTe$_{2}$~since it undergoes a superconducting transition at $T_c \sim 1.7$~K~\cite{Guggenheim1961}. Furthermore, PdTe$_{2}$~ is classified as a type-II Dirac semimetal, as uncovered by angle-resolved photoemission spectroscopy and \textit{ab initio} electronic structure calculations~\cite{Bahramy2018,Liu2015a,Fei2017,Noh2017,Clark2017}. A type-II Dirac semimetal is characterized by a Dirac cone with a tilt parameter $k > 1$ leading to broken Lorentz invariance~\cite{Soluyanov2015}. It is predicted that for Dirac semimetals with $k \approx 1$, meaning close to the topological transition at $k = 1$, superconductivity is generally of the second type (type-II)~\cite{Rosenstein2018}. For $k > 1$, superconductivity becomes of the first kind (type-I). Interestingly, PdTe$_{2}$~\cite{Leng2017,Salis2021} is a type-I superconductor and based on its $T_c$ Shapiro \textit{et al}.~\cite{Shapiro2018} estimated $k \approx 2$. In view of the effect topology has on superconductivity in these systems, it is of interest to investigate whether the superconductivity type can be altered by, for instance, doping. \\

Superconductivity in PdTe$_{2}$~has been explored in great detail. Type-I superconductivity was uncovered with help of magnetic and transport measurements on single crystals~\cite{Leng2017}. The intermediate state, a hallmark of type-I behavior, was observed through the dc magnetization curves and the differential paramagnetic effect in the ac susceptibility data. Here, a bulk critical field $B_c(0) = 13.6$~mT was determined in conjunction with a surface critical field $B_{c}^S(0) = 34.9$~mT. Moreover, the temperature dependence of the surface superconductivity did not follow the Saint-James $-$ de Gennes model~\cite{Saint-James&deGennes1963}. Peculiarly, from resistance measurements a critical field $B_c^R(0) = 0.32$~T was deduced. Weak-coupling conventional superconductivity in PdTe$_{2}$~was demonstrated via measurements of the heat capacity~\cite{Amit2018,Salis2021}, penetration depth~\cite{Teknowijoyo2018,Salis2018}, scanning tunneling microscopy and spectroscopy (STM/STS)~\cite{Clark2017,Das2018,Sirohi2019}, and side junction tunneling spectroscopy~\cite{Voerman2019}. Superconductivity is partly attributed to a van Hove singularity situated at $\sim 30$~meV above the Fermi level~\cite{Kim2018,vanHeumen2019}.

On the other hand, a mixed type-I and type-II superconducting state was concluded from STM/STS~\cite{Das2018,Sirohi2019} and point contact spectroscopy (PCS)~\cite{Le2019} measurements. In a magnetic field a range of critical fields was observed at the surface, which was explained by spatially separated type-I and type-II regions. However, later muon spin rotation measurements~\cite{Leng2019} and scanning squid magnetometry~\cite{Garcia-Campos2021} provide solid evidence for bulk type-I superconductivity probed on the microscopic and macroscopic scale, respectively. Finally, evidence for bulk type-I superconductivity was attained through heat capacity measurements by demonstrating the presence of latent heat~\cite{Salis2021}. Measurements under hydrostatic pressure show that superconductivity is still present at 5.5~GPa~\cite{Furue2021} and remains of the first kind at least till 2.5~GPa~\cite{leng2019p}. \\

Substitution or doping studies using PdTe$_{2}$~are scarce. Kudo \textit{et al}.~\cite{Kudo2016} examined Pd substitution in AuTe$_2$ by preparing a series of Au$_{1-x}$Pd$_{x}$Te$_2$ samples. Bulk superconductivity emerges at $x \approx 0.55$ with $T_c \approx 4.0$ K as evidenced by heat capacity measurements. At lower $x$-values the Te-Te dimer connections stabilize a monoclinic crystal structure in which superconductivity is absent~\cite{Kudo2016}. The strong-coupled nature of superconductivity near $x \approx 0.55$ is attributed to a large density of states (DOS) at the Fermi level. Further increasing the Pd content results in weak coupling superconductivity with lower transition temperatures, as expected from approaching the stoichiometric end compound PdTe$_{2}$. Ryu investigated Cu doping in PdTe$_{2}$~by preparing a series of Cu$_x$PdTe$_2$ samples~\cite{Ryu2015}. Optimal doping was found near $x=0.05$ with bulk superconductivity at $T_c \approx 2.6 $ K~\cite{Ryu2015,Hooda2018}. The increase of $T_c$  is attributed to an increase in the DOS at the Fermi level due to the hybridization of Te-\textit{p} and Cu-\textit{d} orbitals along the $c$-axis, effectively reducing the 2D-nature of this layered material. This is in-line with the Cu atoms being intercalated in the Van der Waals gaps. STM/STS measurements provide evidence that Cu$_{0.05}$PdTe$_2$ is a homogeneous type-II superconductor~\cite{Vasdev2019}. This change, compared to the STM/STS data on PdTe$_{2}$~\cite{Das2018,Sirohi2019} that revealed a mixed type-I/II behavior, is explained by Cu intercalation inducing disorder. This effectively reduces the electron mean free path $l_e$ and the coherence length $\xi$, thus increasing the Ginzburg Landau (GL) parameter $\kappa = \frac{\lambda}{\xi}$ to larger than the $1/\sqrt{2}$ threshold for type-I behavior.

Here we report the results of a doping study, directed to intentionally induce disorder in PdTe$_{2}$~by substituting Pd by iso-electronic Pt. We have prepared Pd$_{1-x}$Pt$_x$Te$_2$ crystals with nominal doping concentrations $x=0.05$ and $x=0.10$. Sample characterization by energy dispersive x-ray spectroscopy (EDX) revealed that Pt did not dissolve homogeneously in the crystals. Notably, small crystals cut from the nominal  $x=0.10$ batch appeared to have $x=0.09$, or the non stoichiometric composition Pd$_{0.97}$Pt$_{<0.004}$Te$_{2.03}$. Transport, magnetic and heat capacity measurements demonstrate a transition from type-I to type-II superconducting behavior upon increasing disorder.

\section{Experimental}
PdTe$_{2}$~crystallizes in the trigonal CdI$_2$ structure (space group $P\bar{3}m1$). Two single-crystalline batches Pd$_{1-x}$Pt$_x$Te$_2$ were prepared with $x=0.05$ and $x=0.10$ using a modified Bridgman technique~\cite{Lyons1976}. The same technique was previously used to prepare PdTe$_{2}$~single crystals~\cite{Leng2017}. Small flat crystals were cut from the prepared batches by a scalpel. The crystals have an area of $2 \times 3$~mm$^2$ and a thickness of about 0.3~mm. Scanning electron microscopy with energy dispersive x-ray spectroscopy (SEM/EDX) was carried out with help of a Hitachi table top microscope TM3000. For details of the SEM/EDX results we refer to the Supplemental Material file~\cite{Supp}. SEM micrographs taken on cut crystals and other sample pieces revealed the final composition can deviate from the nominal one and that Pt did not dissolve in the same amount in all pieces. In fact for the cut crystal with a nominal Pt content of 5 at.\% no Pt was detected. This crystal has a stoichiometric composition with a Pd:Te ratio of 1:2 (the error in these numbers is 1\%). Transport, ac susceptibility and heat capacity measurements were carried out on this sample, which we labeled \#ptnom5. For the experiments on the 10 at.\%Pt concentration we used two crystals. One sample had a composition close to the nominal $x=0.10$ composition Pd$_{0.91}$Pt$_{0.09}$Te$_2$. This sample, labeled \#ptnom10res, was used for transport experiments only. EDX on the second sample showed a small Te excess and a very small Pt content ($<$~0.4\%). Its composition is Pd$_{0.97}$Pt$_{<0.004}$Te$_{2.03}$. This sample was used for transport,  ac susceptibility and heat capacity measurements and it is labeled \#ptnom10. We remark that the EDX determined compositions above each yield the average over a large part of the sample surface and are thus representative for the specific sample. The experimental results on the doped samples are compared with previous resistance, ac susceptibility and heat capacity data taken on a crystal with the stoichiometric 1:2 composition to within 0.5\% as determined by EDX~\cite{Leng2017,Salis2021}. In the following this sample is labelled \#pdte2.

Resistance measurements were performed using the standard four point method in a Quantum Design Physical Property Measurement System (PPMS) down to 2.0~K. Data at lower temperatures were collected in a 3-He refrigerator (Heliox, Oxford Instruments) down to 0.3~K using a low frequency (16 Hz) ac-resistance bridge (Linear Research LR700). The ac susceptibility was measured in the Heliox with a custom-made coil set. Data were also taken with the LR700 bridge, operated at a driving field of 0.026~mT.  The heat capacity was measured using the dual slope thermal relaxation calorimetry technique~\cite{Stewart1983}, using a home-built set-up~\cite{Salis2021}, where each data point is the average of four dual slope measurements. The increase in temperature $\Delta T$ in the measurement of the heat capacity is always in between 1$\%$ and 1.6$\%$ of the bath temperature of the particular measurement. In the ac susceptibility and specific heat experiment the dc magnetic field was applied in the $ab$-plane. The demagnetization factor of the crystals is $N\simeq0.1$, which implies the intermediate state is formed between $(1-N)H_c \simeq 0.9H_c$ and $H_c$ in the case of type-I superconductivity. The resistance and ac susceptibility measurements in field have been carried out by applying the field above $T_c$ and subsequently cooling in field, while the specific heat data in field were taken after zero field cooling and then applying the field.

\section{Results}

\begin{figure}[t]
    \centering
    \includegraphics[width = 8.5cm]{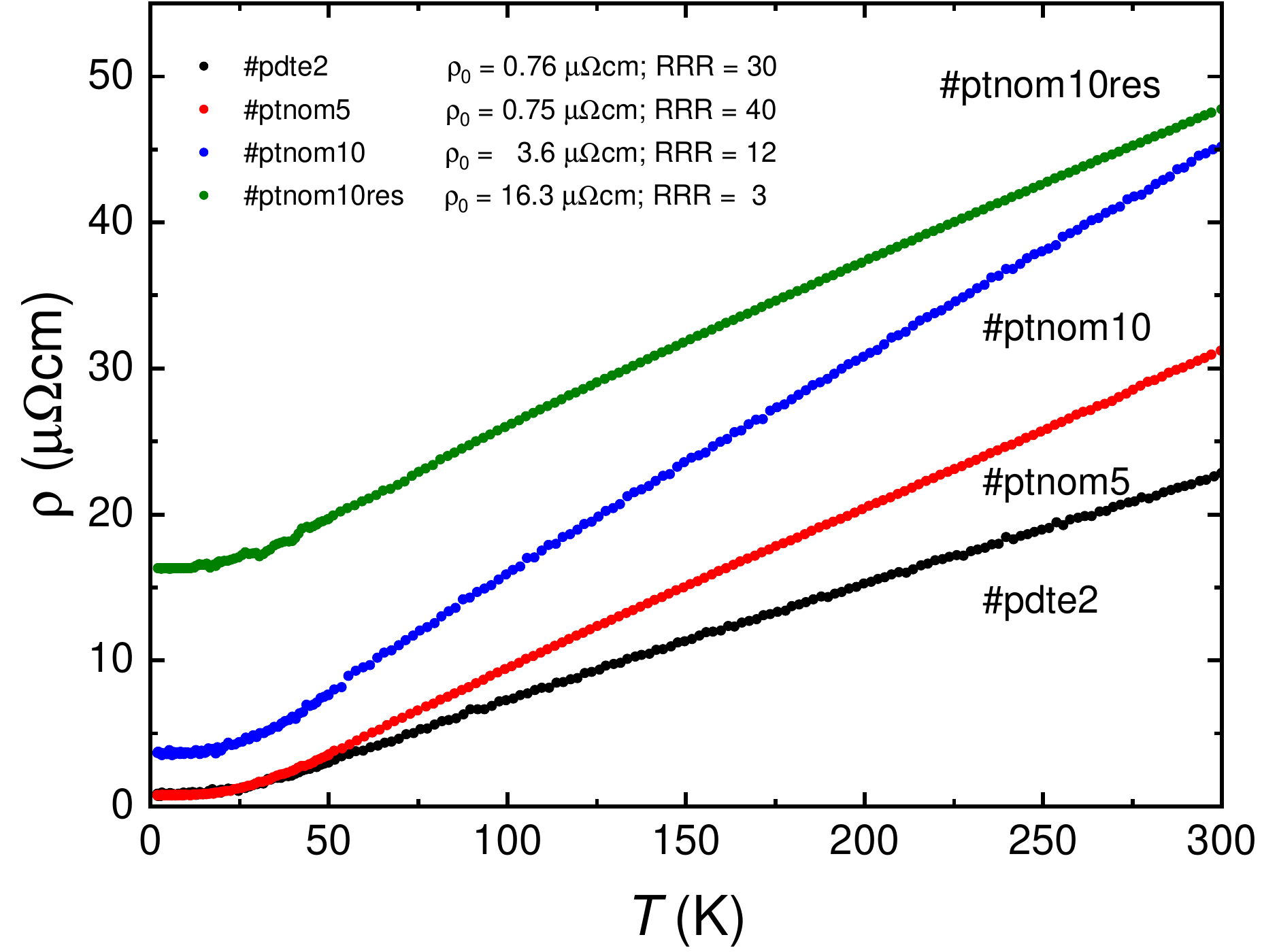}
    \caption{Temperature dependence of the resistivity of crystals \#ptnom5 (red circles), \#ptnom10 (blue circles) and \#ptnom10res (green circles). The data for \#pdte2 (black circles) are taken from Ref.~\cite{Leng2017}.}
    \label{fig:p300}
\end{figure}

The resistivity of samples \#ptnom5, \#ptnom10 and \#ptnom10res in the temperature range 2-300~K is shown in figure \ref{fig:p300}, where we have also traced the data for crystal \#pdte2 reported in Ref.~\cite{Leng2017}. The curves for \#ptnom5 and \#pdte2 are very similar with a residual resistivity value, $\rho_0$, taken at 2~K, of 0.75 and 0.76~$\mu\Omega$cm, respectively. This is in agreement with both samples having the same stoichiometric 1:2 composition. The residual resistance ratio, RRR= $\rho(300 $K$)/\rho_0$, amounts to 40 and 30, respectively. For the non-stoichiometric sample \#ptnom10 $\rho_0$ has increased to 3.6~$\mu\Omega$cm and RRR = 12. The $\rho_0$-value of the substituted sample \#ptnom10res is considerably higher as expected, and equals 16.3~$\mu\Omega$cm. Its RRR is 3.

\begin{figure}[t]
    \centering
    \includegraphics[width=8cm]{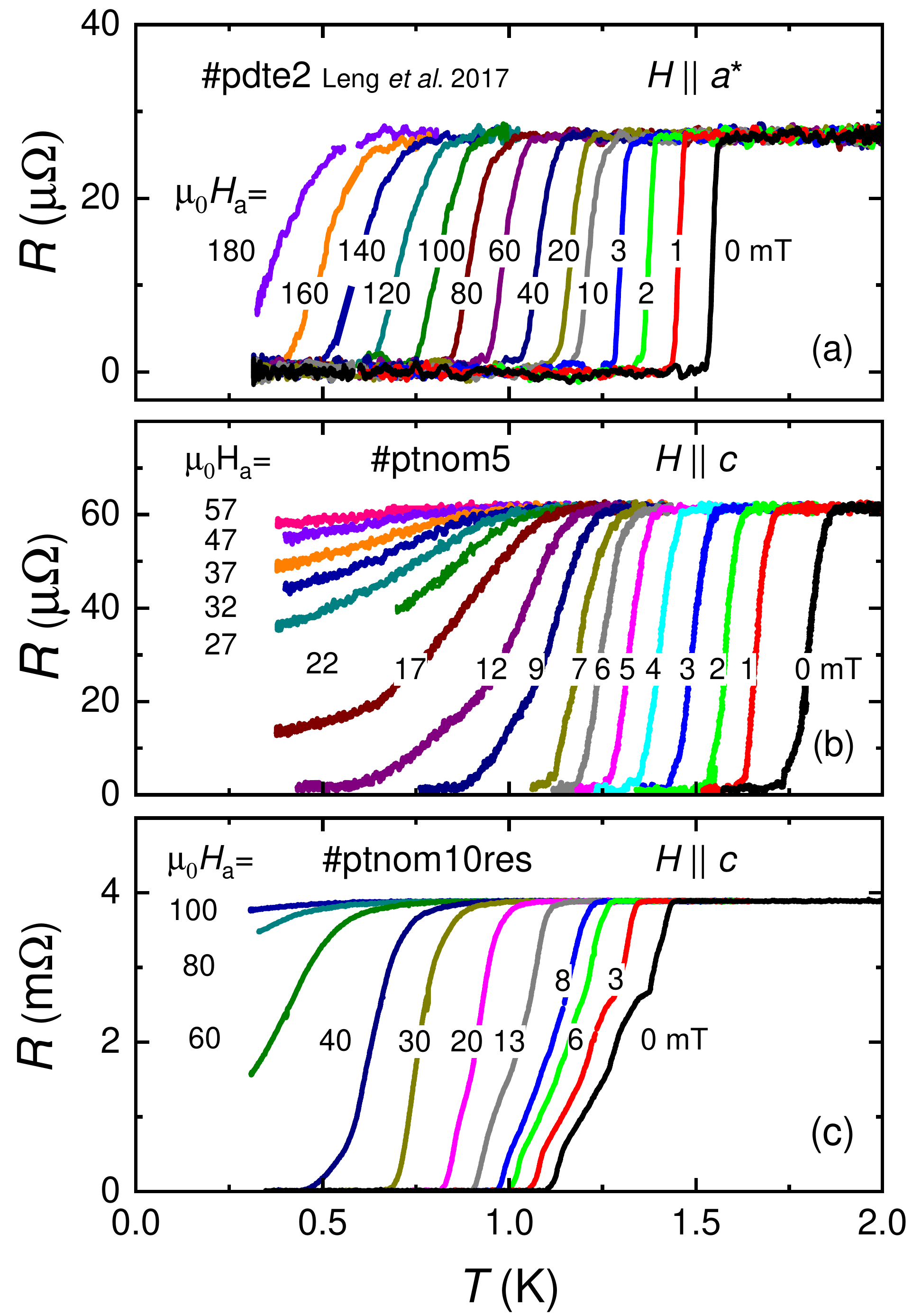}
    \caption{Resistance as a function of temperature around the superconducting transition for crystal \#ptnom5 (panel (b)) and \#ptnom10res (panel (c)) in zero field (black curves) and small applied fields, $\mu_0H_a$, as indicated. The data in panel (a) for \#pdte2 are taken from Ref.~\cite{Leng2017}. }
    \label{fig:res}
\end{figure}

\begin{figure}[t]
    \centering
    \includegraphics[width = 8 cm]{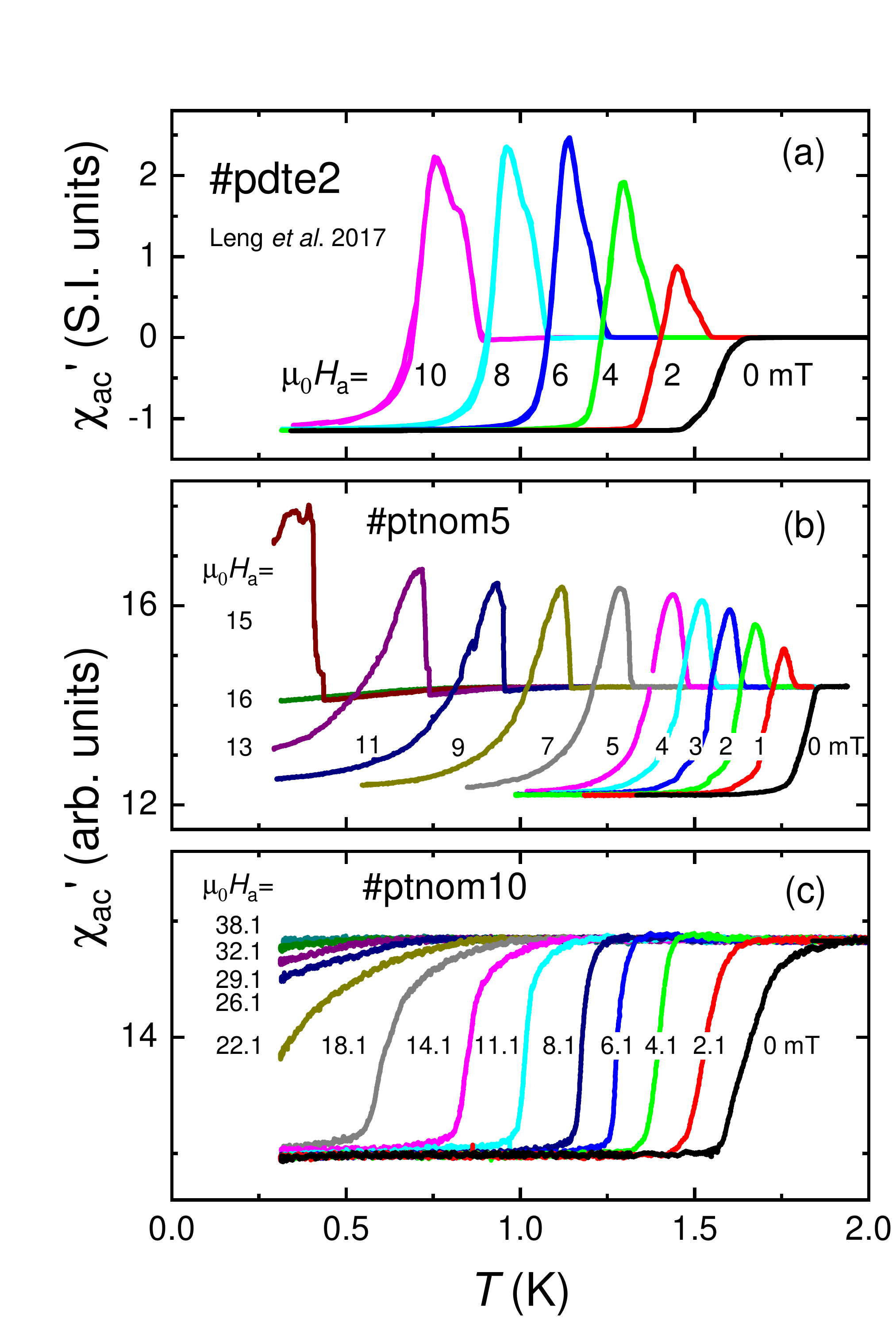}
    \caption{Ac susceptibility of crystals \#ptnom5 (panel (b)) and \#ptnom10 (panel (c)) measured in zero field (black curves) and small applied dc fields as indicated. The field is applied in the $ab$-plane. The data of \#pdte2 are taken from Ref.~\cite{Leng2017}. Note the ac driving field applied to take the data in panels (b) and (c) is a factor 10 smaller than in panel (a).}
    \label{fig:ac}
\end{figure}

\begin{figure}[t]
    \centering
    \includegraphics[width= 8cm]{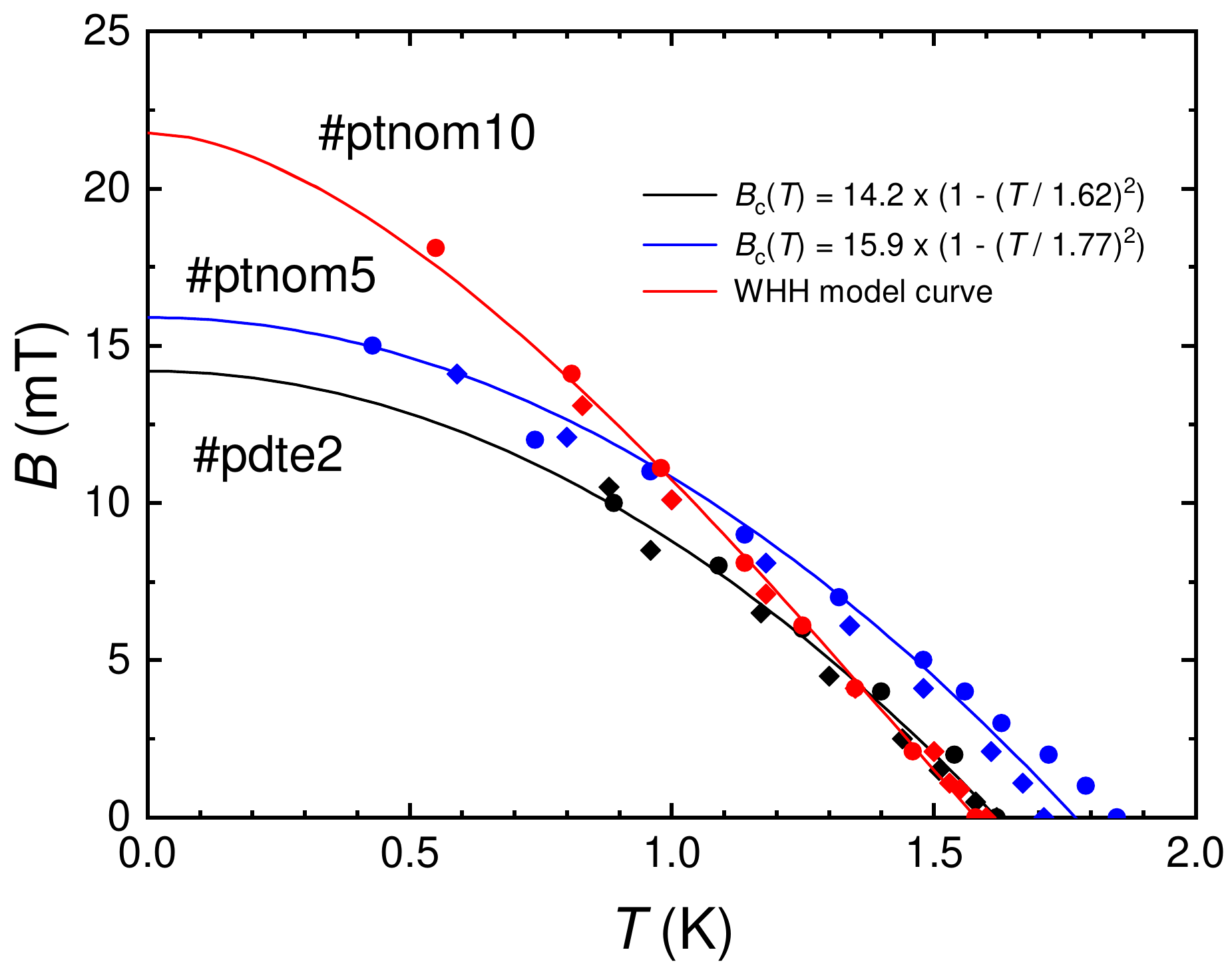}
    \caption{Critical field $B_c(T)$ of crystal \#pdte2 and \#ptnom5 and upper critical field $B_{c2}(T)$ of crystal \#ptnom10 extracted from the specific heat (closed squares) and ac susceptibility (closed circles) data. }
    \label{fig:Bc}
\end{figure}

The resistance as a function of temperature around the superconducting transition in zero field and applied magnetic fields of crystals \#ptnom5 and \#ptnom10res is depicted in figure \ref{fig:res}. Again, the data for \#pdte2, shown in panel (a), are taken from Ref.~\cite{Leng2017}. The critical temperature in zero field, $T_c(0)$, here defined by the onset of the transition, is 1.87~K and 1.56~K for the stoichiometric samples \#ptnom5 and \#pdte2, respectively. Surprisingly, the higher $T_c$ and RRR for \#ptnom5 indicate it has a somewhat higher purity than sample \#pdte2. For the substituted sample the superconducting transition shows several steps and $T_c$ is lower. It ranges from 1.44 to 1.12~K. In a magnetic field superconductivity is rapidly suppressed. The data in panels (b) and (c) of figure \ref{fig:res} show these crystals also have superconducting resistance paths in fields above the critical field $B_c(0)$ determined by ac susceptibility and heat capacity (see below and figure \ref{fig:Bc}). The $B_c^R(0)$ values that can be deduced are however not as large as the value $B_c^R(0) \approx $ 0.3 T for $ H$ $\|~c$ reported for PdTe$_{2}$ (see figure~S6 in the Supplemental Material file of Ref.~\cite{Leng2017}).

In figure \ref{fig:ac} we show the in-phase component of the ac susceptibility, $\chi_{ac}'$, in arbitrary units measured on crystals \#ptnom5 and \#ptnom10 in the temperature range 0.3-2.0~K. Again the data are compared with those of \#pdte2 (data in S.I. units taken from Ref.~\cite{Leng2017}). The onset $T_c$ values are 1.64~K and 1.85~K for \#pdte2 and \#ptnom5 and compare well to the values determined above from the resistivity. The onset $T_c$ value for \#ptnom10 is 1.91~K, but the transition is rather broad (the width is 0.3~K) with a slow decrease below $T_c$. The resistance of this sample was only measured in the PPMS down to 2.0~K. The RRR-value of 12 tells us the disorder is enhanced, which is also reflected in the broad transition. The $\chi_{ac}'(T)$ data measured in applied magnetic fields for \#pdte2 and \#ptnom5 show pronounced peaks below $T_c$ that are due to the differential paramagnetic effect (DPE). The DPE is due to the positive $dM/dH$ ($M$ is the magnetization) in the intermediate state~\cite{Hein1961}. The intermediate phase is due to the sample shape and is present when the demagnetization factor, $N$, is finite. Observation of red a DPE that largely exceeds the Meissner signal can therefore be used as solid proof for type-I superconductivity. Most importantly, the DPE is absent for crystal \#ptnom10, which provides the first piece of evidence it is a type-II superconductor.

\begin{figure}[t]
    \centering
    \includegraphics[width= 8cm]{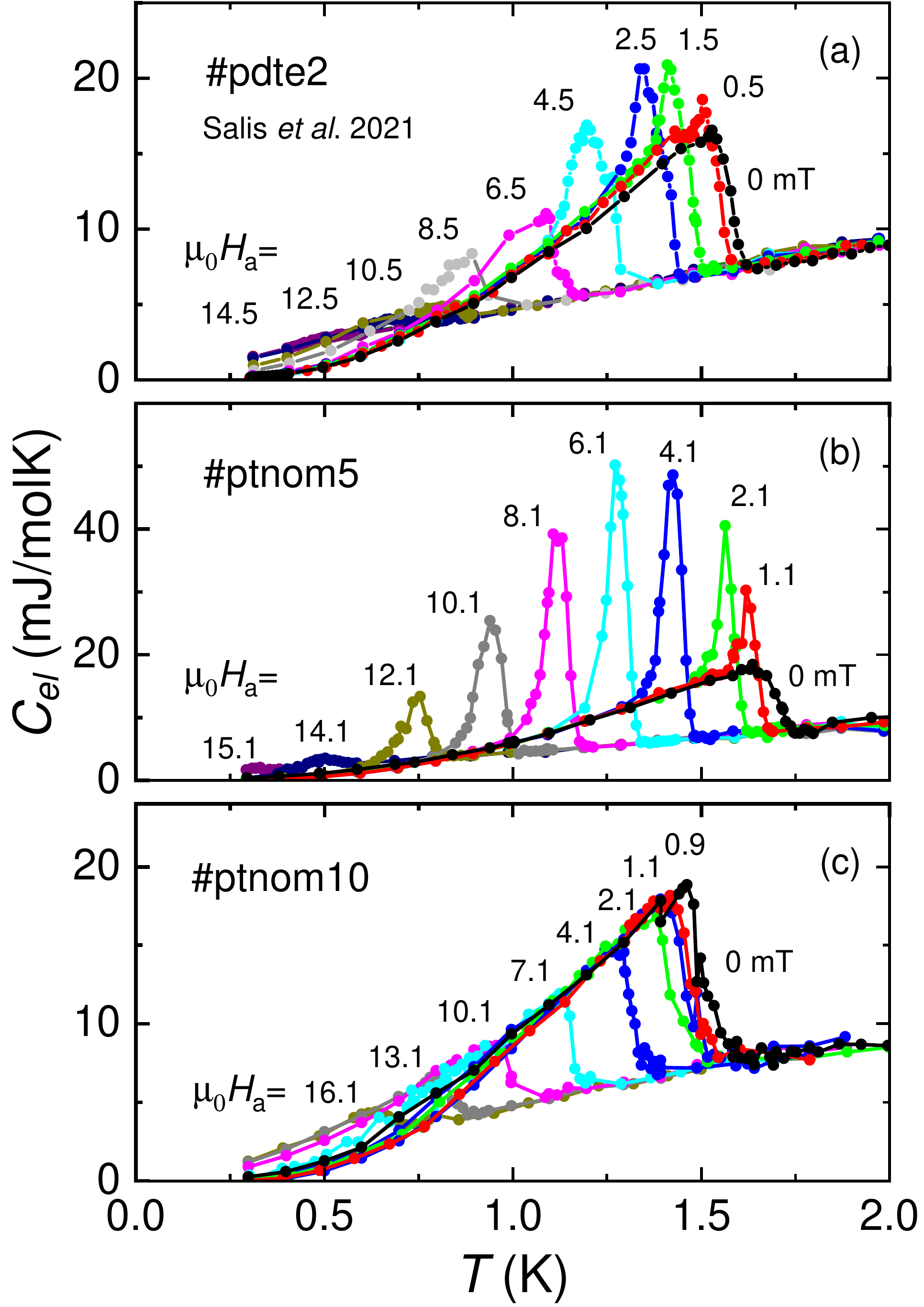}
    \caption{Electronic specific heat, $C_{el}$, of crystals \#ptnom5 (panel (b)) and \#ptnom10 (panel (c)) measured in zero field (black curves) and small applied dc fields as indicated. The field is directed in the $ab$-plane. The data of \#pdte2 are taken from Ref.~\cite{Salis2021}.}
    \label{fig:cel}
\end{figure}

In figure \ref{fig:cel} we show the electronic specific heat, $C_{el}$, of crystals \#ptnom5 and \#ptnom10 in the temperature range 0.3-2.0~K. The $C_{el}(T)$-curves are obtained by subtracting the phononic contribution from the measured $C$ in the standard way, \textit{i.e.} by using the relation $C = \gamma T + \beta T^3$, where $\gamma$ is the Sommerfeld coefficient and $\beta$ the phononic coefficient. The data are compared with $C_{el}$ of PdTe$_{2}$~reported in Ref.~\cite{Salis2021} (panel (a) of figure \ref{fig:cel}). This PdTe$_2$ crystal was cut from the same batch as the samples studied in Ref.~\cite{Leng2017} and we also label it \#pdte2. The onset $T_c$ values of crystals \#pdte2 and \#ptnom5 are 1.62~K and 1.75~K and compare well to the values determined above. The onset $T_c = 1.60$~K for \#ptnom10 is however lower than the value 1.91~K determined by $\chi_{ac}'(T)$.

The $\gamma$-values of the three crystals in panel (a), (b) and (c) of figure \ref{fig:cel} amount to 4.4, 4.5 and 4.7~mJ/molK$^2$ and the $\beta$-values are 0.7, 1.1 and 1.0~mJ/molK$^4$, respectively. These $\gamma$-values are very similar, which indicates the density of states near the Fermi level is not affected much by doping. The $\beta$-values do show some variation, which is not correlated with the amount of disorder, and likely related to an experimental uncertainty because of the small temperature interval in which $\beta$ is obtained. To examine the strength of the electron-phonon coupling, the step size $\Delta C |_{T_c}$ is analysed using the BCS relation $\Delta C |_{T_c}/ \gamma T_c =1.43$, where $T_c$ is the superconducting transition temperature, here taken as the onset of superconductivity. For crystal \#pdte2 a ratio $\Delta C |_{T_c} / \gamma T_c = 1.42$ is found~\cite{Salis2021}, which is close to the textbook value of 1.43 for a weakly coupled BCS superconductor. For crystal \#ptnom5 a ratio of 1.41 is found, which presents a minute change from the textbook value. However, for crystal \#ptnom10 we determine a ratio of 1.48, suggesting that superconductivity is slightly more than weakly coupled.

Next we discuss the electronic specific heat measured in applied magnetic fields (figure \ref{fig:cel}). Distinguishing between type-I and type-II superconductivity via heat capacity can be achieved by observing the presence or absence of latent heat. The extra energy necessary to facilitate a first order phase transition is reflected in the heat capacity as an increased value of $C$ at the transition. A type-I superconductor has a second order phase transition in zero field, but a first order one in field. While for a type-II superconductor the transition remains second order in an applied field. The excess $C_{el}$ above the standard BCS heat capacity in panel (a) provided solid thermodynamic evidence PdTe$_2$ is a type-I superconductor~\cite{Salis2021}. Surprisingly, for crystal \#ptnom5 (panel b) the excess specific heat becomes more pronounced as illustrated by the sharp peaks below $T_c(B)$. Thus the contribution of the latent heat to $C_{el}$ is much larger, which indicates the transition has a stronger first order character than observed for crystal \#pdte2. On the other hand, for crystal \#ptnom10 the data in panel (c) show latent heat is absent, which provides the second piece of evidence of type-II superconductivity, in-line with the $\chi_{ac}'$-data.

Finally, we trace the temperature variation of the critical field, $B_c(T)$, extracted from the ac susceptibility (figure \ref{fig:ac}) and  specific heat data (figure \ref{fig:cel}). The $B-T$ phase diagram is reported in figure \ref{fig:Bc}. For crystals \#pdte2 and \#ptnom5 we identify $T_c(B)$ by the onset in $C_{el}$ and the onset of the DPE in $\chi_{ac}'(T)$. $B_c(T)$ follows the standard BCS quadratic temperature variation $B_c(T) = B_c(0)[1-(T/T_c)^2]$, with $B_c(0) = 14.2$~mT and $T_c$ = 1.63~K for \#ptpde2~\cite{Leng2017,Salis2021}, and $B_c(0) = 15.9$~mT and $T_c$ = 1.77~K for \#ptnom5. For crystal \#ptnom10 the transition in $\chi_{ac}'(T)$ is rather broad. Here we identify $T_c$ by the onset temperature in $C_{el}$, which corresponds to the temperature where the magnetic transition is complete in $\chi_{ac}'(T)$. The $B-T$ phase-line provides further evidence for type-II superconductivity. It compares well to the Werthamer-Helfand-Hohenberg (WHH) model curve~\cite{Werthamer1966} for an orbital-limited weak-coupling spin-singlet superconductor with an upper critical field $B_{c2}(0) = 21.8$~mT.

\section{Discussion}

From the sample preparation side our goal was to prepare Pd$_{1-x}$Pt$_x$Te$_2$ crystals with $x=0.05$ and $x=0.10$. The SEM/EDX micrographs showed that Pt did not dissolve as expected in these crystals and that the single-crystalline batches are inhomogeneous. Crystals cut from the nominal $x=0.05$ batch appeared to be undoped and have the 1:2 stoichiometry. From the nominal $x=0.10$ batch we managed to obtain a crystal with $x=0.09$, and a non-stoichiometric crystal Pd$_{0.97}$Pt$_{<0.004}$Te$_{2.03}$. Specific heat and ac-susceptibility measurements on this last crystal \#ptnom10 demonstrated we could make a doping-induced transition to type-II superconductivity.

To observe type-II superconductivity the disorder should be large enough such that the threshold $\kappa = 1/\sqrt{2}$ can be overcome. The effect of controlled non-magnetic disorder on the normal and superconducting properties of PdTe$_{2}$~was recently studied by electron irradiation by Timmons \textit{et al}.~\cite{Timmons2020}. The residual resistivity was found to increase from a pristine crystal value of 0.6~$\mu\Omega$cm to 2.4~$\mu\Omega$cm for a irradiation dose of 2.4~C/cm$^2$, while at the same time $T_c$ decreased from 1.76~K to 1.65~K as identified by reaching the zero resistance state $R=0$. Assuming a linear relation between $\rho_0$ and $T_c$, $T_c$ decreases at a rate of 0.046~K/$\mu\Omega$cm. With this rate we estimate for crystal \#ptnom10res ($\Delta \rho_0 = 15.5~\mu\Omega$cm) $T_c =0.9$~K, which compares favorably to the measured $T_c = 1.1$~K ($R=0$), given the crude approximation. In this electron irradiation work no discussion was made whether disorder is strong enough to induce type-II behavior.

For crystal \#ptnom10 the coherence length $\xi$ can be calculated from the relation $B_{c2}(0) = \Phi_0 /2\pi\xi^2$, where $\Phi_0$ is the flux quantum. From figure \ref{fig:Bc} we determine $B_{c2}(0) = 21.8$~mT and obtain $\xi = 123$~nm. The coherence length can be related to the electron mean free path, $l_e$, via Pippard's relation $1/\xi = 1/\xi_0 + 1/l_e$, where $\xi_0$ is the intrinsic coherence length given by the  BCS value~\cite{Tinkham1996}. With $\xi_0 = 1.8~\mu$m~\cite{Salis2018} and $\xi =~123$~nm we obtain $\l_e= 132$~nm. As expected, this value is reduced compared to $\l_e= 531$~nm calculated from the residual resistivity value $\rho_0 = 0.76~\mu\Omega$cm~\cite{Salis2018} of nominally pure PdTe$_{2}$. Reversely, using the experimental value $\rho_0 = 3.6~\mu\Omega$cm (figure \ref{fig:p300}) we calculate $\l_e= 112$~nm for crystal \#ptnom10, which is close to the value $\l_e= 132$~nm derived from Pippard's relation.

Next we calculate $\kappa=\lambda/\xi$ of crystal \#ptnom10. In their controlled disorder study Timmons \textit{et al}.~\cite{Timmons2020} measured the penetration depth and found that upon increasing the disorder $\lambda$ stays nearly constant~\cite{Timmons2020} at a value of 220~nm. This is in-line with the minute change in the $\gamma$-value reported above. With $\xi =123$~nm we calculate $\kappa =1.8$, which is in agreement with superconductivity being of the second kind. For crystals \#pdte2 and \#ptnom5 we calculate $\kappa \simeq $~0.5-0.6~\cite{Leng2017}. Here $\xi \simeq $~440-370~nm is estimated from the GL relation $\xi=\Phi_0/(2\sqrt 2 \pi B_c \lambda_L)$~\cite{Tinkham1996}, where $\lambda_L \propto (m_e/n_s)^{1/2}$ is the London penetration depth with $m_e$ the effective electron mass and $n_s$ the superfluid density.

Another way to provide an estimate of $\kappa$ of crystal \#ptnom10 is from the GL relation $\kappa = B_{c2}/\sqrt{2}B_c$. The thermodynamic critical field, $B_c$, can be determined from the specific heat by the relation $\Delta C |_{T_c} = 4 B_c(0)^2 /\mu_0T_c$~\cite{Poole2007}, where $C$ is in units of J/m$^3$. From $\Delta C |_{T_c}$ in figure \ref{fig:cel} (panel (c)) we calculate $B_c(0) = 11.1$~mT. We remark this value is close to the calculated value $B_c(0) = 12.6$~mT reported for PdTe$_{2}$~\cite{Leng2017}. Using $B_c(0)=11.1$~mT and $B_{c2}= 21.8$~mT in the expression above, we calculate $\kappa = 1.4$, which is similar to the  value of 1.8 directly estimated from the ratio $\lambda / \xi$.

We remark that for Type-I superconductivity $B_{c}(0)$ can also be obtained from the latent heat with help of the Clausius-Clapeyron relation. We calculate $B_c(0)$ = 11.2 mT and 11.1~mT for \#pdte2 and \#ptnom5, respectively, in good agreement with the values obtained from $\Delta C |_{T_c}$ in zero field~\cite{Supp}.

Our results are of relevance for the observation of a mixed type-I and type-II superconducting state in PdTe$_2$ probed by surface sensitive techniques~\cite{Das2018,Sirohi2019,Le2019}. Our doping study shows that nominal pure PdTe$_2$ crystals can already be close to the type-I/II border. Using the value $\lambda =230$~nm~\cite{Timmons2020}, we calculate $\xi = 310$~nm at the threshold value $\kappa=1/\sqrt{2}$. This implies $l_e$ should be smaller than 375~nm for type-II superconductivity, or $\rho_0 > 1.4~\mu\Omega$cm. From the resistivity graph reported in Ref.~\cite{Das2018} we deduce $\rho_0 \simeq 1~\mu\Omega$cm, which indeed is not far from the type-I/II border. Thus it is plausible inhomogeneities give rise to the mixed phase observation reported in Refs.~\cite{Das2018,Sirohi2019,Le2019}.

An unsolved aspect of superconductivity in PdTe$_2$ is the observation of surface superconductivity detected in the screening signal $\chi_{ac}'(T)$ measured in small applied dc fields~\cite{Leng2017,leng2019p}. The extracted surface critical field $B_{c}^S(0) = 34.9$~mT exceeds the value predicted by the Saint-James $-$ de Gennes model~\cite{Saint-James&deGennes1963} $B_{c3} = 2.39 \times \kappa B_c = 16.3$~mT. Recently, the GL model at the superconducting-insulator boundary was revisited~\cite{Samoilenka2021} and it was shown that $T_c$ and the third critical field $B_{c3}$ can be enhanced to exceed the Saint-James $-$ de Gennes value, which is worthy to explore further. On the other hand, it is tempting to attribute the surface superconductivity in PdTe$_2$ to superconductivity of the topological surface state detected by ARPES~\cite{Bahramy2018,Liu2015a,Fei2017,Noh2017,Clark2017}. We remark that the $\chi_{ac}'(T)$ data for the doped crystals, reported in figure \ref{fig:ac} panel (b) and (c), also show superconducting screening signals above the $B_c(0)$ and $B_{c2}(0)$-values reported in figure \ref{fig:Bc}. Likewise, the resistance traces in figure \ref{fig:res} reveal $B_c^R(0)$ is similarly enhanced. These screening signals of enhanced superconductivity are however not as pronounced as reported for PdTe$_2$ in Ref.~\cite{Leng2017}. Nonetheless, the robustness of superconducting screening signals above $B_c(0)$ or $B_{c2}(0)$ upon doping, as well as under high pressure~\cite{leng2019p}, calls for further experiments.

\section{Conclusion}

The Dirac semimetal PdTe$_{2}$~is a type-I superconductor with $T_c = 1.7$~K. We have carried out a doping study directed to intentionally increase the disorder and induce type-II superconductivity. Two single-crystalline batches Pd$_{1-x}$Pt$_x$Te$_2$ have been prepared with nominal doping concentrations $x=0.05$ and $x=0.10$. Sample characterization by energy dispersive x-ray spectroscopy (EDX) on small crystals cut from the batches revealed that Pt did not dissolve homogeneously in the crystals. In fact the nominal $x=0.05$ crystal appeared to be undoped and have the stoichiometric 1:2 composition. From the nominal $x=0.10$ batch we obtained a small single crystal with $x=0.09$, as well as a crystal with the non stoichiometric composition Pd$_{0.97}$Pt$_{<0.004}$Te$_{2.03}$. The presence of the differential paramagnetic effect in the ac susceptibility and latent heat in the heat capacity demonstrate the nominal $x=0.05$ crystal is a type-I superconductor, just like PdTe$_{2}$. The absence of these effects for Pd$_{0.97}$Pt$_{<0.004}$Te$_{2.03}$ revealed it is a type-II superconductor with an upper critical field $B_{c2} = 21.8$~mT. The analysis of $B_{c2}$ and resistance data using Pippard's model convincingly show PdTe$_{2}$~can be turned into a superconductor of the second kind when the residual resistivity $\rho_0 > 1.4~\mu\Omega$cm.

\

Acknowledgement: This work is part of the Projectruimte programme with project number 680-91-109, which is financed by the Dutch Research Council (NWO).

\pagebreak
\widetext
\begin{center}
\textbf{\large Supplemental Material: Disorder induced transition from type-I to type-II superconductivity in the Dirac semimetal PdTe$_{2}$ }
\end{center}

\setcounter{equation}{0}
\setcounter{figure}{0}
\setcounter{table}{0}
\makeatletter
\renewcommand{\theequation}{S\arabic{equation}}
\renewcommand{\thefigure}{S\arabic{figure}}
\renewcommand{\bibnumfmt}[1]{[S#1]}
\renewcommand{\citenumfont}[1]{S#1}
\vspace{3cm}

\large{Content} \\

\normalsize
1. SEM/EDX mapping experimental and results\\

\hspace{1cm} Fig.S1  SEM/EDX mapping of crystal \#ptnom5\\

\hspace{1cm} Fig.S2  SEM/EDX mapping of crystal \#ptnom10\\

\hspace{1cm} Fig.S3  SEM/EDX mapping of crystal \#ptnom10res\\

2. Latent heat and Clausius-Clapeyron relation\\
\newpage
\
\begin{centering}
\large{1. SEM/EDX mapping: experimental}\\
\end{centering}
\normalsize
Scanning electron microscopy with energy dispersive x-ray spectroscopy (SEM/EDX) was carried out with help of a Hitachi table top microscope TM3000. The acceleration voltage in all measurements is 15 kV. Of each single-crystalline boule prepared with a certain nominal Pt content small thin crystals were isolated with typical size 2  $\times$ 3 mm$^2$. On each of these crystals we have investigated the composition by EDX in several areas of typically 200 $\times$ 200 $\mu$m$^2$. Overall the composition in the sleected and measured crystals was found to be homogenous. The table below gives the labels of the doped crystals, the nominal Pt content and the EDX determined composition.

\begin{table}[h]
\begin{tabular}{|c|c|c|}
\hline
Crystal & Nominal Pt cotent & EDX compostion\\\hline
\#ptnom5& 5 at. $\%$ & PdTe$_2$ \\ \hline
\#ptnom10 & 10 at. $\%$& Pd$_{0.97}$Pt$_{<0.004}$Te$_{2.03}$ \\ \hline
\#ptnom10res & 10 at.$\%$ & Pd$_{0.91}$Pt$_{0.09}$Te$_2$ \\ \hline

\end{tabular}
\end{table}

Typical SEM/EDX results for crystals \#ptnom5, \#ptnom10 and \#ptnom10res are given in Fig. S1, Fig. S2 and Fig. S3, respectively. Shown are:\\

\textbf{Top panel}: The SEM spectrum in cps/eV (counts per second per electron-volt) with Pd, Pt and Te peaks labelled. \\

\textbf{Middle panel}: Table with quantitative results of the composition analysis.\\

\textbf{Lower left panel}: SEM picture of the crystal with scanned area for the composition analysis indicated.\\

\textbf{Lower right panel}: Pd, Pt and Te element distribution in the scanned area.

\newpage
\centering
\large{\textbf{Figure S1 SEM/EDX mapping of crystal \#ptnom5}}
\begin{figure}[H]
    \centering
    \includegraphics[width=15cm]{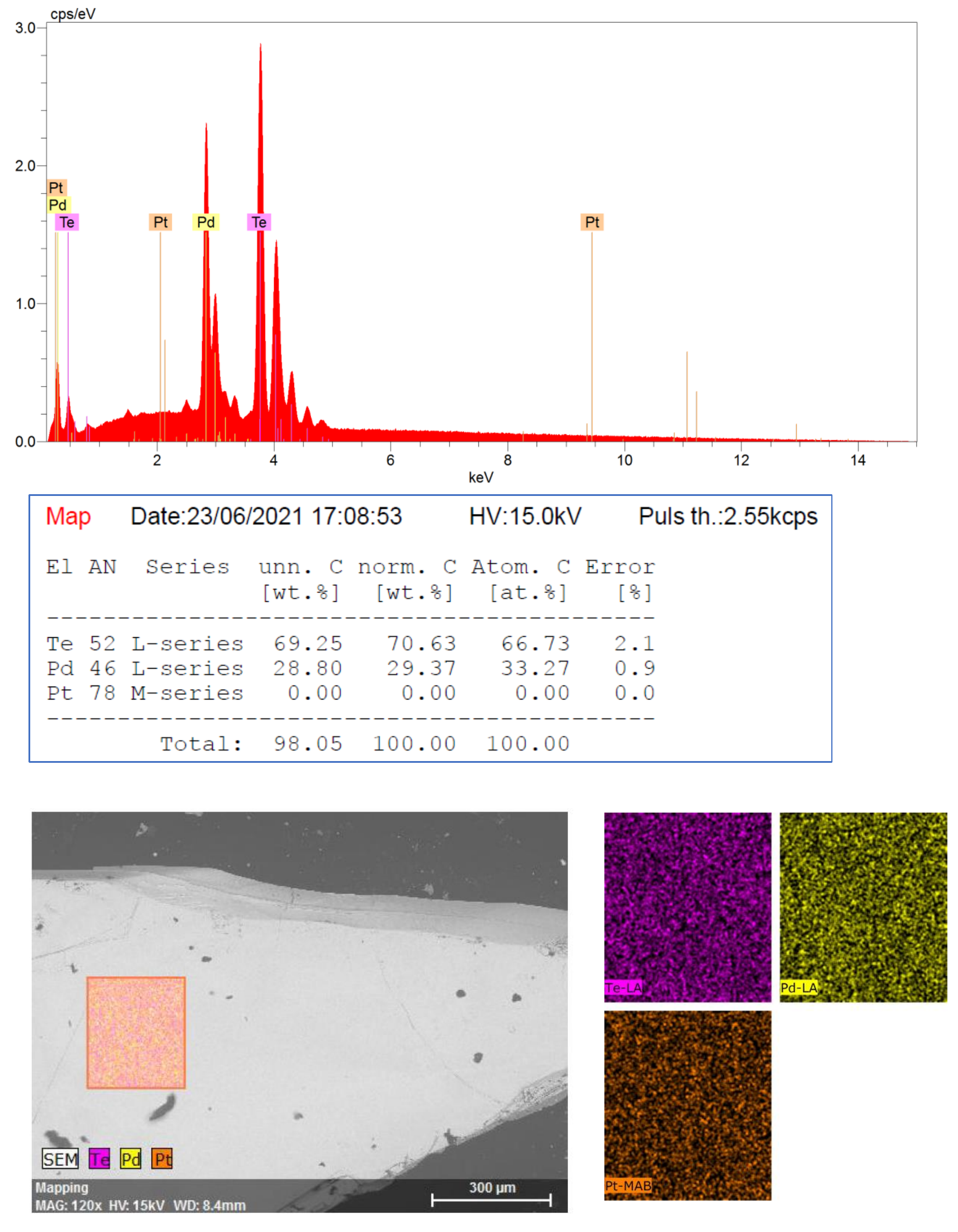}
    \label{fig:s1}
\end{figure}
\newpage
\large{\textbf{Figure S2 SEM/EDX mapping of crystal \#ptnom10}}
\begin{figure}[H]
    \centering
    \includegraphics[width=15cm]{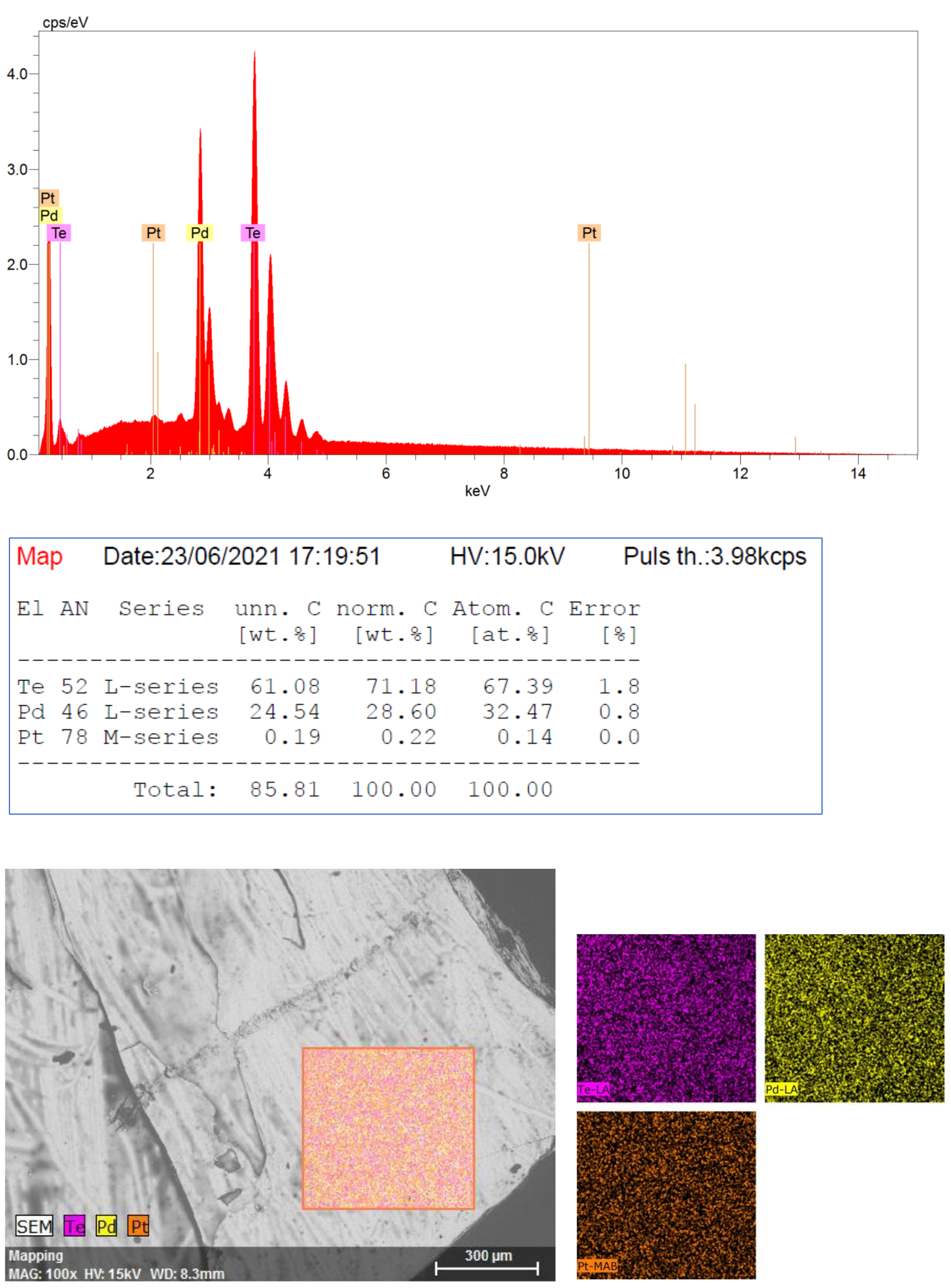}
    \label{fig:s2}
\end{figure}
\newpage
\large{\textbf{Figure S3 SEM/EDX mapping of crystal \#ptnom10res}}
\begin{figure}[H]
    \centering
    \includegraphics[width=15cm]{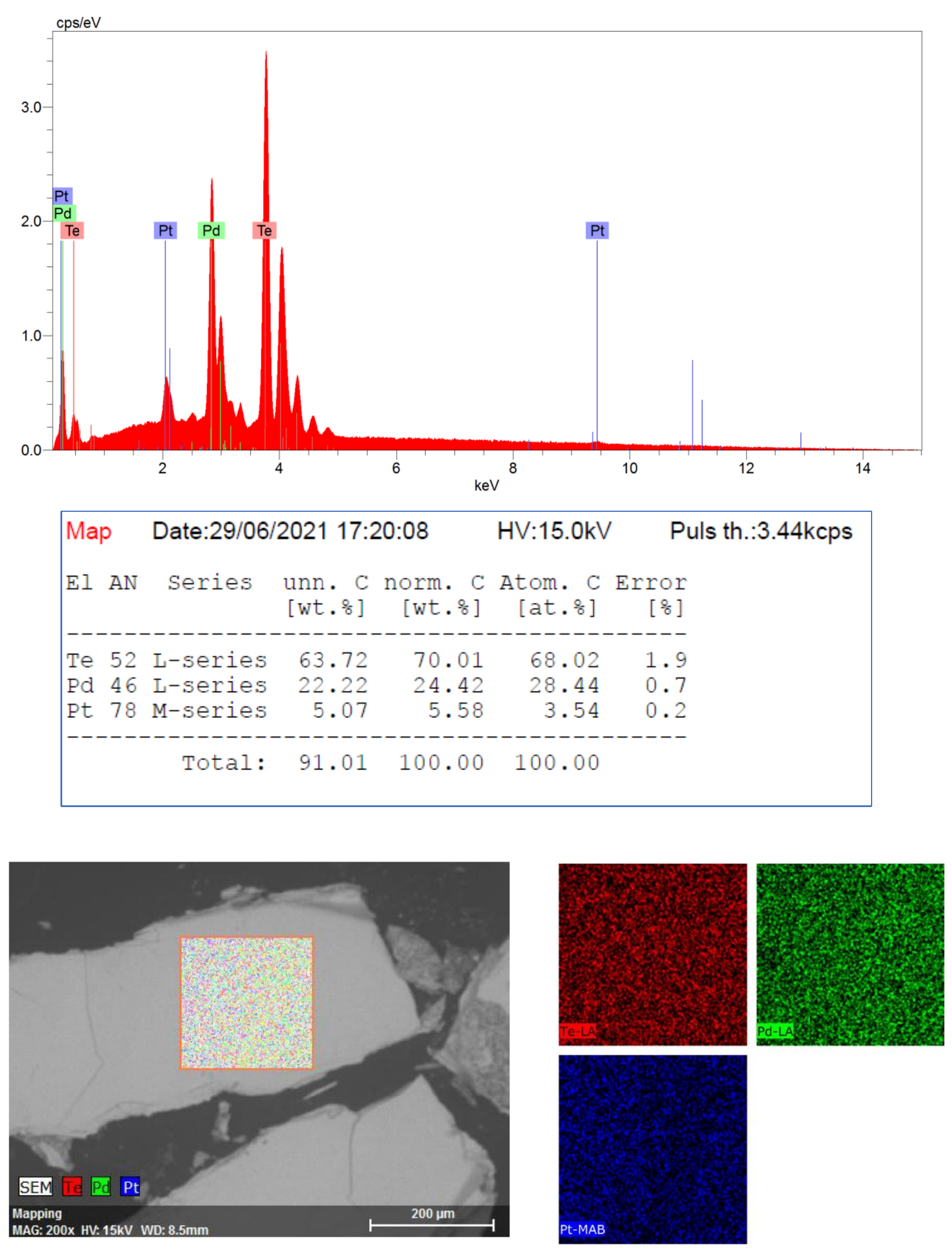}
    \label{fig:s3}
\end{figure}

\pagebreak
\begin{centering}
\large{2. Latent heat and Clausius-Clapeyron relation}\\
\end{centering}
\normalsize
As mentioned in the Discussion section in the manuscript an estimate for the thermodynamic critical field $B_c(0)$ can be obtained by evaluating the relation $\Delta C = 4B_c(0)^2 / \mu_0 T_c$, where $\Delta C$ is the step at $T_c$ in the specific heat at $B = 0$. Alternatively, for a type-I superconductor in a magnetic field $B_c(0)$ can be calculated from the latent heat, $L$. Using the Clausius-Clapeyron equation the entropy change is given by $\Delta S = L(T)/T = - 2\mu_0 B_c(T) \frac{dB_c(T)}{dT}$, with $B_c(T) = B_c(0)(1-(T/T_c)^2)$. $\Delta S$ was calculated by graphically integrating the $C/T$ data in field and subtracting the data taken at $B = 0$. This procedure is carried out for all applied field values, and from $\Delta S(B)$ we obtain $B_c(0)$ as a fit parameter. For crystals \#pdte2 and \#ptnom5, that exhibit type-I superconductivity, we obtain $B_c(0) = 11.2$ mT and 11.1 mT, respectively. We remark these values are close to the $B_c(0)$ values evaluated at $B = 0$. On the other hand, the $B_c(0)$-values derived from $\Delta C$ and $\Delta S$ are smaller than the value derived directly from the experiment. In the table below we compare the $B_c(0)$ values obtained in different ways. 
 
\begin{centering}
\begin{table}[h]
\begin{tabular}{|c|c|c|c|}
\hline
Crystal&$B_c(0)$ from $\Delta C$ &$B_c(0)$ from $\Delta S$ &$B_c(0)$ experiment\\
\hline
\#pdte2 &10.9 mT& 11.2 mT & 14.2 mT \\
\hline
\#ptnom5 &11.8 mT& 11.1 mT & 15.9 mT \\
\hline
\#ptnom10 &11.1 mT& - & - \\
\hline
\end{tabular}
\end{table}
\end{centering}

\begin{thebibliography}{40}%
\makeatletter
\providecommand \@ifxundefined [1]{%
 \@ifx{#1\undefined}
}%
\providecommand \@ifnum [1]{%
 \ifnum #1\expandafter \@firstoftwo
 \else \expandafter \@secondoftwo
 \fi
}%
\providecommand \@ifx [1]{%
 \ifx #1\expandafter \@firstoftwo
 \else \expandafter \@secondoftwo
 \fi
}%
\providecommand \natexlab [1]{#1}%
\providecommand \enquote  [1]{``#1''}%
\providecommand \bibnamefont  [1]{#1}%
\providecommand \bibfnamefont [1]{#1}%
\providecommand \citenamefont [1]{#1}%
\providecommand \href@noop [0]{\@secondoftwo}%
\providecommand \href [0]{\begingroup \@sanitize@url \@href}%
\providecommand \@href[1]{\@@startlink{#1}\@@href}%
\providecommand \@@href[1]{\endgroup#1\@@endlink}%
\providecommand \@sanitize@url [0]{\catcode `\\12\catcode `\$12\catcode
  `\&12\catcode `\#12\catcode `\^12\catcode `\_12\catcode `\%12\relax}%
\providecommand \@@startlink[1]{}%
\providecommand \@@endlink[0]{}%
\providecommand \url  [0]{\begingroup\@sanitize@url \@url }%
\providecommand \@url [1]{\endgroup\@href {#1}{\urlprefix }}%
\providecommand \urlprefix  [0]{URL }%
\providecommand \Eprint [0]{\href }%
\providecommand \doibase [0]{http://dx.doi.org/}%
\providecommand \selectlanguage [0]{\@gobble}%
\providecommand \bibinfo  [0]{\@secondoftwo}%
\providecommand \bibfield  [0]{\@secondoftwo}%
\providecommand \translation [1]{[#1]}%
\providecommand \BibitemOpen [0]{}%
\providecommand \bibitemStop [0]{}%
\providecommand \bibitemNoStop [0]{.\EOS\space}%
\providecommand \EOS [0]{\spacefactor3000\relax}%
\providecommand \BibitemShut  [1]{\csname bibitem#1\endcsname}%
\let\auto@bib@innerbib\@empty
\bibitem [{\citenamefont {Soluyanov}\ \emph {et~al.}(2015)\citenamefont
  {Soluyanov}, \citenamefont {Gresch}, \citenamefont {Wang}, \citenamefont
  {Wu}, \citenamefont {Troyer}, \citenamefont {Dai},\ and\ \citenamefont
  {Bernevig}}]{Soluyanov2015}%
  \BibitemOpen
  \bibfield  {author} {\bibinfo {author} {\bibfnamefont {A.~A.}\ \bibnamefont
  {Soluyanov}}, \bibinfo {author} {\bibfnamefont {D.}~\bibnamefont {Gresch}},
  \bibinfo {author} {\bibfnamefont {Z.}~\bibnamefont {Wang}}, \bibinfo {author}
  {\bibfnamefont {Q.}~\bibnamefont {Wu}}, \bibinfo {author} {\bibfnamefont
  {M.}~\bibnamefont {Troyer}}, \bibinfo {author} {\bibfnamefont
  {X.}~\bibnamefont {Dai}}, \ and\ \bibinfo {author} {\bibfnamefont
  {B.}~\bibnamefont {Bernevig}},\ }\href@noop {} {\bibfield  {journal}
  {\bibinfo  {journal} {Nature}\ }\textbf {\bibinfo {volume} {257}},\ \bibinfo
  {pages} {495} (\bibinfo {year} {2015})}\BibitemShut {NoStop}%
\bibitem [{\citenamefont {Huang}\ \emph {et~al.}(2016)\citenamefont {Huang},
  \citenamefont {Zhou},\ and\ \citenamefont {Duan}}]{Huang2016}%
  \BibitemOpen
  \bibfield  {author} {\bibinfo {author} {\bibfnamefont {H.}~\bibnamefont
  {Huang}}, \bibinfo {author} {\bibfnamefont {S.}~\bibnamefont {Zhou}}, \ and\
  \bibinfo {author} {\bibfnamefont {W.}~\bibnamefont {Duan}},\ }\href {\doibase
  10.1103/PhysRevB.94.121117} {\bibfield  {journal} {\bibinfo  {journal} {Phys.
  Rev. B}\ }\textbf {\bibinfo {volume} {94}},\ \bibinfo {pages} {121117}
  (\bibinfo {year} {2016})}\BibitemShut {NoStop}%
\bibitem [{\citenamefont {Yan}\ \emph {et~al.}(2017)\citenamefont {Yan},
  \citenamefont {Huang}, \citenamefont {Zhang}, \citenamefont {Wang},
  \citenamefont {Yao}, \citenamefont {Deng}, \citenamefont {Wan}, \citenamefont
  {Zhang}, \citenamefont {Arita}, \citenamefont {Yang}, \citenamefont {Sun},
  \citenamefont {Yao}, \citenamefont {Wu}, \citenamefont {Fan}, \citenamefont
  {Duan},\ and\ \citenamefont {Zhou}}]{Yan2017}%
  \BibitemOpen
  \bibfield  {author} {\bibinfo {author} {\bibfnamefont {M.}~\bibnamefont
  {Yan}}, \bibinfo {author} {\bibfnamefont {H.}~\bibnamefont {Huang}}, \bibinfo
  {author} {\bibfnamefont {K.}~\bibnamefont {Zhang}}, \bibinfo {author}
  {\bibfnamefont {E.}~\bibnamefont {Wang}}, \bibinfo {author} {\bibfnamefont
  {W.}~\bibnamefont {Yao}}, \bibinfo {author} {\bibfnamefont {K.}~\bibnamefont
  {Deng}}, \bibinfo {author} {\bibfnamefont {G.}~\bibnamefont {Wan}}, \bibinfo
  {author} {\bibfnamefont {H.}~\bibnamefont {Zhang}}, \bibinfo {author}
  {\bibfnamefont {M.}~\bibnamefont {Arita}}, \bibinfo {author} {\bibfnamefont
  {H.}~\bibnamefont {Yang}}, \bibinfo {author} {\bibfnamefont {Z.}~\bibnamefont
  {Sun}}, \bibinfo {author} {\bibfnamefont {H.}~\bibnamefont {Yao}}, \bibinfo
  {author} {\bibfnamefont {Y.}~\bibnamefont {Wu}}, \bibinfo {author}
  {\bibfnamefont {S.}~\bibnamefont {Fan}}, \bibinfo {author} {\bibfnamefont
  {W.}~\bibnamefont {Duan}}, \ and\ \bibinfo {author} {\bibfnamefont
  {S.}~\bibnamefont {Zhou}},\ }\href@noop {} {\bibfield  {journal} {\bibinfo
  {journal} {Nature Comm.}\ }\textbf {\bibinfo {volume} {8}},\ \bibinfo {pages}
  {257} (\bibinfo {year} {2017})}\BibitemShut {NoStop}%
\bibitem [{\citenamefont {Bahramy}\ \emph {et~al.}(2018)\citenamefont
  {Bahramy}, \citenamefont {Clark}, \citenamefont {Yang}, \citenamefont {Feng},
  \citenamefont {Bawden}, \citenamefont {Riley}, \citenamefont {Markovic},
  \citenamefont {Mazzola}, \citenamefont {Sunko}, \citenamefont {Biswas},
  \citenamefont {Cooil}, \citenamefont {Jorge}, \citenamefont {Wells},
  \citenamefont {Leandersson}, \citenamefont {Balasubramanian}, \citenamefont
  {Fujii}, \citenamefont {Vobornik}, \citenamefont {Rault}, \citenamefont
  {Kim}, \citenamefont {Hoesch}, \citenamefont {Okawa}, \citenamefont
  {Asakawa}, \citenamefont {Sasagawa}, \citenamefont {Eknapakul}, \citenamefont
  {Meevasana},\ and\ \citenamefont {King}}]{Bahramy2018}%
  \BibitemOpen
  \bibfield  {author} {\bibinfo {author} {\bibfnamefont {M.~S.}\ \bibnamefont
  {Bahramy}}, \bibinfo {author} {\bibfnamefont {O.~J.}\ \bibnamefont {Clark}},
  \bibinfo {author} {\bibfnamefont {B.-J.}\ \bibnamefont {Yang}}, \bibinfo
  {author} {\bibfnamefont {J.}~\bibnamefont {Feng}}, \bibinfo {author}
  {\bibfnamefont {L.}~\bibnamefont {Bawden}}, \bibinfo {author} {\bibfnamefont
  {J.~M.}\ \bibnamefont {Riley}}, \bibinfo {author} {\bibfnamefont
  {I.}~\bibnamefont {Markovic}}, \bibinfo {author} {\bibfnamefont
  {F.}~\bibnamefont {Mazzola}}, \bibinfo {author} {\bibfnamefont
  {V.}~\bibnamefont {Sunko}}, \bibinfo {author} {\bibfnamefont
  {D.}~\bibnamefont {Biswas}}, \bibinfo {author} {\bibfnamefont {S.~P.}\
  \bibnamefont {Cooil}}, \bibinfo {author} {\bibfnamefont {M.}~\bibnamefont
  {Jorge}}, \bibinfo {author} {\bibfnamefont {J.~W.}\ \bibnamefont {Wells}},
  \bibinfo {author} {\bibfnamefont {M.}~\bibnamefont {Leandersson}}, \bibinfo
  {author} {\bibfnamefont {T.}~\bibnamefont {Balasubramanian}}, \bibinfo
  {author} {\bibfnamefont {J.}~\bibnamefont {Fujii}}, \bibinfo {author}
  {\bibfnamefont {I.}~\bibnamefont {Vobornik}}, \bibinfo {author}
  {\bibfnamefont {J.~E.}\ \bibnamefont {Rault}}, \bibinfo {author}
  {\bibfnamefont {T.~K.}\ \bibnamefont {Kim}}, \bibinfo {author} {\bibfnamefont
  {M.}~\bibnamefont {Hoesch}}, \bibinfo {author} {\bibfnamefont
  {K.}~\bibnamefont {Okawa}}, \bibinfo {author} {\bibfnamefont
  {M.}~\bibnamefont {Asakawa}}, \bibinfo {author} {\bibfnamefont
  {T.}~\bibnamefont {Sasagawa}}, \bibinfo {author} {\bibfnamefont
  {T.}~\bibnamefont {Eknapakul}}, \bibinfo {author} {\bibfnamefont
  {W.}~\bibnamefont {Meevasana}}, \ and\ \bibinfo {author} {\bibfnamefont
  {P.~D.~C.}\ \bibnamefont {King}},\ }\href {\doibase 10.1038/nmat5031}
  {\bibfield  {journal} {\bibinfo  {journal} {Nature Mat.}\ }\textbf {\bibinfo
  {volume} {17}},\ \bibinfo {pages} {21} (\bibinfo {year} {2018})}\BibitemShut
  {NoStop}%
\bibitem [{\citenamefont {Guggenheim}\ \emph {et~al.}(1961)\citenamefont
  {Guggenheim}, \citenamefont {Hulliger},\ and\ \citenamefont
  {M\"{u}ller}}]{Guggenheim1961}%
  \BibitemOpen
  \bibfield  {author} {\bibinfo {author} {\bibfnamefont {J.}~\bibnamefont
  {Guggenheim}}, \bibinfo {author} {\bibfnamefont {F.}~\bibnamefont
  {Hulliger}}, \ and\ \bibinfo {author} {\bibfnamefont {J.}~\bibnamefont
  {M\"{u}ller}},\ }\href@noop {} {\bibfield  {journal} {\bibinfo  {journal}
  {Helv. Phys. Acta}\ }\textbf {\bibinfo {volume} {34}},\ \bibinfo {pages}
  {408} (\bibinfo {year} {1961})}\BibitemShut {NoStop}%
\bibitem [{\citenamefont {Yan}\ \emph {et~al.}(2015)\citenamefont {Yan},
  \citenamefont {Jian-Zhou}, \citenamefont {Li}, \citenamefont {Cheng-Tian},
  \citenamefont {Ai-Ji}, \citenamefont {Cheng}, \citenamefont {Ying},
  \citenamefont {Yu}, \citenamefont {Shao-Long}, \citenamefont {Lin},
  \citenamefont {Guo-Dong}, \citenamefont {Xiao-Li}, \citenamefont {Jun},
  \citenamefont {Chuang-Tian}, \citenamefont {Zu-Yan}, \citenamefont
  {Hong-Ming}, \citenamefont {Xi}, \citenamefont {Zhong},\ and\ \citenamefont
  {Xing-Jiang}}]{Liu2015a}%
  \BibitemOpen
  \bibfield  {author} {\bibinfo {author} {\bibfnamefont {L.}~\bibnamefont
  {Yan}}, \bibinfo {author} {\bibfnamefont {Z.}~\bibnamefont {Jian-Zhou}},
  \bibinfo {author} {\bibfnamefont {Y.}~\bibnamefont {Li}}, \bibinfo {author}
  {\bibfnamefont {L.}~\bibnamefont {Cheng-Tian}}, \bibinfo {author}
  {\bibfnamefont {L.}~\bibnamefont {Ai-Ji}}, \bibinfo {author} {\bibfnamefont
  {H.}~\bibnamefont {Cheng}}, \bibinfo {author} {\bibfnamefont
  {D.}~\bibnamefont {Ying}}, \bibinfo {author} {\bibfnamefont {X.}~\bibnamefont
  {Yu}}, \bibinfo {author} {\bibfnamefont {H.}~\bibnamefont {Shao-Long}},
  \bibinfo {author} {\bibfnamefont {Z.}~\bibnamefont {Lin}}, \bibinfo {author}
  {\bibfnamefont {L.}~\bibnamefont {Guo-Dong}}, \bibinfo {author}
  {\bibfnamefont {D.}~\bibnamefont {Xiao-Li}}, \bibinfo {author} {\bibfnamefont
  {Z.}~\bibnamefont {Jun}}, \bibinfo {author} {\bibfnamefont {C.}~\bibnamefont
  {Chuang-Tian}}, \bibinfo {author} {\bibfnamefont {X.}~\bibnamefont {Zu-Yan}},
  \bibinfo {author} {\bibfnamefont {W.}~\bibnamefont {Hong-Ming}}, \bibinfo
  {author} {\bibfnamefont {D.}~\bibnamefont {Xi}}, \bibinfo {author}
  {\bibfnamefont {F.}~\bibnamefont {Zhong}}, \ and\ \bibinfo {author}
  {\bibfnamefont {Z.}~\bibnamefont {Xing-Jiang}},\ }\href {\doibase
  10.1088/0256-307X/32/6/067303} {\bibfield  {journal} {\bibinfo  {journal}
  {Chin. Phys. Lett.}\ }\textbf {\bibinfo {volume} {32}},\ \bibinfo {eid}
  {067303} (\bibinfo {year} {2015})}\BibitemShut {NoStop}%
\bibitem [{\citenamefont {Fei}\ \emph {et~al.}(2017)\citenamefont {Fei},
  \citenamefont {Bo}, \citenamefont {Wang}, \citenamefont {Wu}, \citenamefont
  {Jiang}, \citenamefont {Fu}, \citenamefont {Gao}, \citenamefont {Zheng},
  \citenamefont {Chen}, \citenamefont {Wang}, \citenamefont {Bu}, \citenamefont
  {Song}, \citenamefont {Wan}, \citenamefont {Wang},\ and\ \citenamefont
  {Wang}}]{Fei2017}%
  \BibitemOpen
  \bibfield  {author} {\bibinfo {author} {\bibfnamefont {F.}~\bibnamefont
  {Fei}}, \bibinfo {author} {\bibfnamefont {X.}~\bibnamefont {Bo}}, \bibinfo
  {author} {\bibfnamefont {R.}~\bibnamefont {Wang}}, \bibinfo {author}
  {\bibfnamefont {B.}~\bibnamefont {Wu}}, \bibinfo {author} {\bibfnamefont
  {J.}~\bibnamefont {Jiang}}, \bibinfo {author} {\bibfnamefont
  {D.}~\bibnamefont {Fu}}, \bibinfo {author} {\bibfnamefont {M.}~\bibnamefont
  {Gao}}, \bibinfo {author} {\bibfnamefont {H.}~\bibnamefont {Zheng}}, \bibinfo
  {author} {\bibfnamefont {Y.}~\bibnamefont {Chen}}, \bibinfo {author}
  {\bibfnamefont {X.}~\bibnamefont {Wang}}, \bibinfo {author} {\bibfnamefont
  {H.}~\bibnamefont {Bu}}, \bibinfo {author} {\bibfnamefont {F.}~\bibnamefont
  {Song}}, \bibinfo {author} {\bibfnamefont {X.}~\bibnamefont {Wan}}, \bibinfo
  {author} {\bibfnamefont {B.}~\bibnamefont {Wang}}, \ and\ \bibinfo {author}
  {\bibfnamefont {G.}~\bibnamefont {Wang}},\ }\href {\doibase
  10.1103/PhysRevB.96.041201} {\bibfield  {journal} {\bibinfo  {journal} {Phys.
  Rev. B}\ }\textbf {\bibinfo {volume} {96}},\ \bibinfo {pages} {041201}
  (\bibinfo {year} {2017})}\BibitemShut {NoStop}%
\bibitem [{\citenamefont {Noh}\ \emph {et~al.}(2017)\citenamefont {Noh},
  \citenamefont {Jeong}, \citenamefont {Cho}, \citenamefont {Kim},
  \citenamefont {Min},\ and\ \citenamefont {Park}}]{Noh2017}%
  \BibitemOpen
  \bibfield  {author} {\bibinfo {author} {\bibfnamefont {H.-J.}\ \bibnamefont
  {Noh}}, \bibinfo {author} {\bibfnamefont {J.}~\bibnamefont {Jeong}}, \bibinfo
  {author} {\bibfnamefont {E.-J.}\ \bibnamefont {Cho}}, \bibinfo {author}
  {\bibfnamefont {K.}~\bibnamefont {Kim}}, \bibinfo {author} {\bibfnamefont
  {B.~I.}\ \bibnamefont {Min}}, \ and\ \bibinfo {author} {\bibfnamefont
  {B.-G.}\ \bibnamefont {Park}},\ }\href {\doibase
  10.1103/PhysRevLett.119.016401} {\bibfield  {journal} {\bibinfo  {journal}
  {Phys. Rev. Lett.}\ }\textbf {\bibinfo {volume} {119}},\ \bibinfo {pages}
  {016401} (\bibinfo {year} {2017})}\BibitemShut {NoStop}%
\bibitem [{\citenamefont {Clark}\ \emph {et~al.}(2018)\citenamefont {Clark},
  \citenamefont {Neat}, \citenamefont {Okawa}, \citenamefont {Bawden},
  \citenamefont {Markovi\ifmmode~\acute{c}\else \'{c}\fi{}}, \citenamefont
  {Mazzola}, \citenamefont {Feng}, \citenamefont {Sunko}, \citenamefont
  {Riley}, \citenamefont {Meevasana}, \citenamefont {Fujii}, \citenamefont
  {Vobornik}, \citenamefont {Kim}, \citenamefont {Hoesch}, \citenamefont
  {Sasagawa}, \citenamefont {Wahl}, \citenamefont {Bahramy},\ and\
  \citenamefont {King}}]{Clark2017}%
  \BibitemOpen
  \bibfield  {author} {\bibinfo {author} {\bibfnamefont {O.~J.}\ \bibnamefont
  {Clark}}, \bibinfo {author} {\bibfnamefont {M.~J.}\ \bibnamefont {Neat}},
  \bibinfo {author} {\bibfnamefont {K.}~\bibnamefont {Okawa}}, \bibinfo
  {author} {\bibfnamefont {L.}~\bibnamefont {Bawden}}, \bibinfo {author}
  {\bibfnamefont {I.}~\bibnamefont {Markovi\ifmmode~\acute{c}\else
  \'{c}\fi{}}}, \bibinfo {author} {\bibfnamefont {F.}~\bibnamefont {Mazzola}},
  \bibinfo {author} {\bibfnamefont {J.}~\bibnamefont {Feng}}, \bibinfo {author}
  {\bibfnamefont {V.}~\bibnamefont {Sunko}}, \bibinfo {author} {\bibfnamefont
  {J.~M.}\ \bibnamefont {Riley}}, \bibinfo {author} {\bibfnamefont
  {W.}~\bibnamefont {Meevasana}}, \bibinfo {author} {\bibfnamefont
  {J.}~\bibnamefont {Fujii}}, \bibinfo {author} {\bibfnamefont
  {I.}~\bibnamefont {Vobornik}}, \bibinfo {author} {\bibfnamefont {T.~K.}\
  \bibnamefont {Kim}}, \bibinfo {author} {\bibfnamefont {M.}~\bibnamefont
  {Hoesch}}, \bibinfo {author} {\bibfnamefont {T.}~\bibnamefont {Sasagawa}},
  \bibinfo {author} {\bibfnamefont {P.}~\bibnamefont {Wahl}}, \bibinfo {author}
  {\bibfnamefont {M.~S.}\ \bibnamefont {Bahramy}}, \ and\ \bibinfo {author}
  {\bibfnamefont {P.~D.~C.}\ \bibnamefont {King}},\ }\href {\doibase
  10.1103/PhysRevLett.120.156401} {\bibfield  {journal} {\bibinfo  {journal}
  {Phys. Rev. Lett.}\ }\textbf {\bibinfo {volume} {120}},\ \bibinfo {pages}
  {156401} (\bibinfo {year} {2018})}\BibitemShut {NoStop}%
\bibitem [{\citenamefont {Rosenstein}\ \emph {et~al.}(2018)\citenamefont
  {Rosenstein}, \citenamefont {Shapiro}, \citenamefont {Li},\ and\
  \citenamefont {Shapiro}}]{Rosenstein2018}%
  \BibitemOpen
  \bibfield  {author} {\bibinfo {author} {\bibfnamefont {B.}~\bibnamefont
  {Rosenstein}}, \bibinfo {author} {\bibfnamefont {B.~Y.}\ \bibnamefont
  {Shapiro}}, \bibinfo {author} {\bibfnamefont {D.}~\bibnamefont {Li}}, \ and\
  \bibinfo {author} {\bibfnamefont {I.}~\bibnamefont {Shapiro}},\ }\href
  {\doibase 10.1103/PhysRevB.97.144510} {\bibfield  {journal} {\bibinfo
  {journal} {Phys. Rev. B}\ }\textbf {\bibinfo {volume} {97}},\ \bibinfo
  {pages} {144510} (\bibinfo {year} {2018})}\BibitemShut {NoStop}%
\bibitem [{\citenamefont {Leng}\ \emph {et~al.}(2017)\citenamefont {Leng},
  \citenamefont {Paulsen}, \citenamefont {Huang},\ and\ \citenamefont
  {de~Visser}}]{Leng2017}%
  \BibitemOpen
  \bibfield  {author} {\bibinfo {author} {\bibfnamefont {H.}~\bibnamefont
  {Leng}}, \bibinfo {author} {\bibfnamefont {C.}~\bibnamefont {Paulsen}},
  \bibinfo {author} {\bibfnamefont {Y.~K.}\ \bibnamefont {Huang}}, \ and\
  \bibinfo {author} {\bibfnamefont {A.}~\bibnamefont {de~Visser}},\ }\href
  {\doibase 10.1103/PhysRevB.96.220506} {\bibfield  {journal} {\bibinfo
  {journal} {Phys. Rev. B}\ }\textbf {\bibinfo {volume} {96}},\ \bibinfo
  {pages} {220506} (\bibinfo {year} {2017})}\BibitemShut {NoStop}%
\bibitem [{\citenamefont {Salis}\ \emph {et~al.}(2021)\citenamefont {Salis},
  \citenamefont {Huang},\ and\ \citenamefont {de~Visser}}]{Salis2021}%
  \BibitemOpen
  \bibfield  {author} {\bibinfo {author} {\bibfnamefont {M.~V.}\ \bibnamefont
  {Salis}}, \bibinfo {author} {\bibfnamefont {Y.~K.}\ \bibnamefont {Huang}}, \
  and\ \bibinfo {author} {\bibfnamefont {A.}~\bibnamefont {de~Visser}},\ }\href
  {\doibase 10.1103/PhysRevB.103.104502} {\bibfield  {journal} {\bibinfo
  {journal} {Phys. Rev. B}\ }\textbf {\bibinfo {volume} {103}},\ \bibinfo
  {pages} {104502} (\bibinfo {year} {2021})}\BibitemShut {NoStop}%
\bibitem [{\citenamefont {Shapiro}\ \emph {et~al.}(2018)\citenamefont
  {Shapiro}, \citenamefont {Shapiro}, \citenamefont {Li},\ and\ \citenamefont
  {Rosenstein}}]{Shapiro2018}%
  \BibitemOpen
  \bibfield  {author} {\bibinfo {author} {\bibfnamefont {B.~Y.}\ \bibnamefont
  {Shapiro}}, \bibinfo {author} {\bibfnamefont {I.}~\bibnamefont {Shapiro}},
  \bibinfo {author} {\bibfnamefont {D.}~\bibnamefont {Li}}, \ and\ \bibinfo
  {author} {\bibfnamefont {B.}~\bibnamefont {Rosenstein}},\ }\href {\doibase
  10.1088/1361-648x/aad305} {\bibfield  {journal} {\bibinfo  {journal} {Journal
  of Physics: Condensed Matter}\ }\textbf {\bibinfo {volume} {30}},\ \bibinfo
  {pages} {335403} (\bibinfo {year} {2018})}\BibitemShut {NoStop}%
\bibitem [{\citenamefont {Saint-James}\ and\ \citenamefont
  {de~Gennes}(1963)}]{Saint-James&deGennes1963}%
  \BibitemOpen
  \bibfield  {author} {\bibinfo {author} {\bibfnamefont {D.}~\bibnamefont
  {Saint-James}}\ and\ \bibinfo {author} {\bibfnamefont {P.~G.}\ \bibnamefont
  {de~Gennes}},\ }\href@noop {} {\bibfield  {journal} {\bibinfo  {journal}
  {Phys. Lett.}\ }\textbf {\bibinfo {volume} {7}},\ \bibinfo {pages} {306}
  (\bibinfo {year} {1963})}\BibitemShut {NoStop}%
\bibitem [{\citenamefont {Amit}\ and\ \citenamefont {Singh}(2018)}]{Amit2018}%
  \BibitemOpen
  \bibfield  {author} {\bibinfo {author} {\bibnamefont {Amit}}\ and\ \bibinfo
  {author} {\bibfnamefont {Y.}~\bibnamefont {Singh}},\ }\href {\doibase
  10.1103/PhysRevB.97.054515} {\bibfield  {journal} {\bibinfo  {journal} {Phys.
  Rev. B}\ }\textbf {\bibinfo {volume} {97}},\ \bibinfo {pages} {054515}
  (\bibinfo {year} {2018})}\BibitemShut {NoStop}%
\bibitem [{\citenamefont {Teknowijoyo}\ \emph {et~al.}(2018)\citenamefont
  {Teknowijoyo}, \citenamefont {Jo}, \citenamefont {Scheurer}, \citenamefont
  {Tanatar}, \citenamefont {Cho}, \citenamefont {Bud'ko}, \citenamefont {Orth},
  \citenamefont {Canfield},\ and\ \citenamefont {Prozorov}}]{Teknowijoyo2018}%
  \BibitemOpen
  \bibfield  {author} {\bibinfo {author} {\bibfnamefont {S.}~\bibnamefont
  {Teknowijoyo}}, \bibinfo {author} {\bibfnamefont {N.~H.}\ \bibnamefont {Jo}},
  \bibinfo {author} {\bibfnamefont {M.~S.}\ \bibnamefont {Scheurer}}, \bibinfo
  {author} {\bibfnamefont {M.~A.}\ \bibnamefont {Tanatar}}, \bibinfo {author}
  {\bibfnamefont {K.}~\bibnamefont {Cho}}, \bibinfo {author} {\bibfnamefont
  {S.~L.}\ \bibnamefont {Bud'ko}}, \bibinfo {author} {\bibfnamefont {P.~P.}\
  \bibnamefont {Orth}}, \bibinfo {author} {\bibfnamefont {P.~C.}\ \bibnamefont
  {Canfield}}, \ and\ \bibinfo {author} {\bibfnamefont {R.}~\bibnamefont
  {Prozorov}},\ }\href {\doibase 10.1103/PhysRevB.98.024508} {\bibfield
  {journal} {\bibinfo  {journal} {Phys. Rev. B}\ }\textbf {\bibinfo {volume}
  {98}},\ \bibinfo {pages} {024508} (\bibinfo {year} {2018})}\BibitemShut
  {NoStop}%
\bibitem [{\citenamefont {Salis}\ \emph {et~al.}(2018)\citenamefont {Salis},
  \citenamefont {Rodi{\`{e}}re}, \citenamefont {Leng}, \citenamefont {Huang},\
  and\ \citenamefont {de~Visser}}]{Salis2018}%
  \BibitemOpen
  \bibfield  {author} {\bibinfo {author} {\bibfnamefont {M.~V.}\ \bibnamefont
  {Salis}}, \bibinfo {author} {\bibfnamefont {P.}~\bibnamefont
  {Rodi{\`{e}}re}}, \bibinfo {author} {\bibfnamefont {H.}~\bibnamefont {Leng}},
  \bibinfo {author} {\bibfnamefont {Y.~K.}\ \bibnamefont {Huang}}, \ and\
  \bibinfo {author} {\bibfnamefont {A.}~\bibnamefont {de~Visser}},\ }\href
  {\doibase 10.1088/1361-648x/aaed31} {\bibfield  {journal} {\bibinfo
  {journal} {Journal of Physics: Condensed Matter}\ }\textbf {\bibinfo {volume}
  {30}},\ \bibinfo {pages} {505602} (\bibinfo {year} {2018})}\BibitemShut
  {NoStop}%
\bibitem [{\citenamefont {Das}\ \emph {et~al.}(2018)\citenamefont {Das},
  \citenamefont {Amit}, \citenamefont {Sirohi}, \citenamefont {Yadav},
  \citenamefont {Gayen}, \citenamefont {Singh},\ and\ \citenamefont
  {Sheet}}]{Das2018}%
  \BibitemOpen
  \bibfield  {author} {\bibinfo {author} {\bibfnamefont {S.}~\bibnamefont
  {Das}}, \bibinfo {author} {\bibnamefont {Amit}}, \bibinfo {author}
  {\bibfnamefont {A.}~\bibnamefont {Sirohi}}, \bibinfo {author} {\bibfnamefont
  {L.}~\bibnamefont {Yadav}}, \bibinfo {author} {\bibfnamefont
  {S.}~\bibnamefont {Gayen}}, \bibinfo {author} {\bibfnamefont
  {Y.}~\bibnamefont {Singh}}, \ and\ \bibinfo {author} {\bibfnamefont
  {G.}~\bibnamefont {Sheet}},\ }\href {\doibase 10.1103/PhysRevB.97.014523}
  {\bibfield  {journal} {\bibinfo  {journal} {Phys. Rev. B}\ }\textbf {\bibinfo
  {volume} {97}},\ \bibinfo {pages} {014523} (\bibinfo {year}
  {2018})}\BibitemShut {NoStop}%
\bibitem [{\citenamefont {Sirohi}\ \emph {et~al.}(2019)\citenamefont {Sirohi},
  \citenamefont {Das}, \citenamefont {Adhikary}, \citenamefont {Chowdhury},
  \citenamefont {Vashist}, \citenamefont {Singh}, \citenamefont {Gayen},
  \citenamefont {Das},\ and\ \citenamefont {Sheet}}]{Sirohi2019}%
  \BibitemOpen
  \bibfield  {author} {\bibinfo {author} {\bibfnamefont {A.}~\bibnamefont
  {Sirohi}}, \bibinfo {author} {\bibfnamefont {S.}~\bibnamefont {Das}},
  \bibinfo {author} {\bibfnamefont {P.}~\bibnamefont {Adhikary}}, \bibinfo
  {author} {\bibfnamefont {R.~R.}\ \bibnamefont {Chowdhury}}, \bibinfo {author}
  {\bibfnamefont {A.}~\bibnamefont {Vashist}}, \bibinfo {author} {\bibfnamefont
  {Y.}~\bibnamefont {Singh}}, \bibinfo {author} {\bibfnamefont
  {S.}~\bibnamefont {Gayen}}, \bibinfo {author} {\bibfnamefont
  {T.}~\bibnamefont {Das}}, \ and\ \bibinfo {author} {\bibfnamefont
  {G.}~\bibnamefont {Sheet}},\ }\href {\doibase 10.1088/1361-648x/aaf49c}
  {\bibfield  {journal} {\bibinfo  {journal} {Journal of Physics: Condensed
  Matter}\ }\textbf {\bibinfo {volume} {31}},\ \bibinfo {pages} {085701}
  (\bibinfo {year} {2019})}\BibitemShut {NoStop}%
\bibitem [{\citenamefont {Voerman}\ \emph {et~al.}(2019)\citenamefont
  {Voerman}, \citenamefont {de~Boer}, \citenamefont {Hashimoto}, \citenamefont
  {Huang}, \citenamefont {Li},\ and\ \citenamefont {Brinkman}}]{Voerman2019}%
  \BibitemOpen
  \bibfield  {author} {\bibinfo {author} {\bibfnamefont {J.~A.}\ \bibnamefont
  {Voerman}}, \bibinfo {author} {\bibfnamefont {J.~C.}\ \bibnamefont
  {de~Boer}}, \bibinfo {author} {\bibfnamefont {T.}~\bibnamefont {Hashimoto}},
  \bibinfo {author} {\bibfnamefont {Y.}~\bibnamefont {Huang}}, \bibinfo
  {author} {\bibfnamefont {C.}~\bibnamefont {Li}}, \ and\ \bibinfo {author}
  {\bibfnamefont {A.}~\bibnamefont {Brinkman}},\ }\href {\doibase
  10.1103/PhysRevB.99.014510} {\bibfield  {journal} {\bibinfo  {journal} {Phys.
  Rev. B}\ }\textbf {\bibinfo {volume} {99}},\ \bibinfo {pages} {014510}
  (\bibinfo {year} {2019})}\BibitemShut {NoStop}%
\bibitem [{\citenamefont {Kim}\ \emph {et~al.}(2018)\citenamefont {Kim},
  \citenamefont {Kim}, \citenamefont {Kim}, \citenamefont {Kim}, \citenamefont
  {Park},\ and\ \citenamefont {Min}}]{Kim2018}%
  \BibitemOpen
  \bibfield  {author} {\bibinfo {author} {\bibfnamefont {K.}~\bibnamefont
  {Kim}}, \bibinfo {author} {\bibfnamefont {S.}~\bibnamefont {Kim}}, \bibinfo
  {author} {\bibfnamefont {J.~S.}\ \bibnamefont {Kim}}, \bibinfo {author}
  {\bibfnamefont {H.}~\bibnamefont {Kim}}, \bibinfo {author} {\bibfnamefont
  {J.-H.}\ \bibnamefont {Park}}, \ and\ \bibinfo {author} {\bibfnamefont
  {B.~I.}\ \bibnamefont {Min}},\ }\href {\doibase 10.1103/PhysRevB.97.165102}
  {\bibfield  {journal} {\bibinfo  {journal} {Phys. Rev. B}\ }\textbf {\bibinfo
  {volume} {97}},\ \bibinfo {pages} {165102} (\bibinfo {year}
  {2018})}\BibitemShut {NoStop}%
\bibitem [{\citenamefont {van Heumen}\ \emph {et~al.}(2019)\citenamefont {van
  Heumen}, \citenamefont {Berben}, \citenamefont {Neubrand},\ and\
  \citenamefont {Huang}}]{vanHeumen2019}%
  \BibitemOpen
  \bibfield  {author} {\bibinfo {author} {\bibfnamefont {E.}~\bibnamefont {van
  Heumen}}, \bibinfo {author} {\bibfnamefont {M.}~\bibnamefont {Berben}},
  \bibinfo {author} {\bibfnamefont {L.}~\bibnamefont {Neubrand}}, \ and\
  \bibinfo {author} {\bibfnamefont {Y.}~\bibnamefont {Huang}},\ }\href
  {\doibase 10.1103/PhysRevMaterials.3.114202} {\bibfield  {journal} {\bibinfo
  {journal} {Phys. Rev. Materials}\ }\textbf {\bibinfo {volume} {3}},\ \bibinfo
  {pages} {114202} (\bibinfo {year} {2019})}\BibitemShut {NoStop}%
\bibitem [{\citenamefont {Le}\ \emph {et~al.}(2019)\citenamefont {Le},
  \citenamefont {Yin}, \citenamefont {Feng}, \citenamefont {Huang},
  \citenamefont {Che}, \citenamefont {Li}, \citenamefont {Shi},\ and\
  \citenamefont {Lu}}]{Le2019}%
  \BibitemOpen
  \bibfield  {author} {\bibinfo {author} {\bibfnamefont {T.}~\bibnamefont
  {Le}}, \bibinfo {author} {\bibfnamefont {L.}~\bibnamefont {Yin}}, \bibinfo
  {author} {\bibfnamefont {Z.}~\bibnamefont {Feng}}, \bibinfo {author}
  {\bibfnamefont {Q.}~\bibnamefont {Huang}}, \bibinfo {author} {\bibfnamefont
  {L.}~\bibnamefont {Che}}, \bibinfo {author} {\bibfnamefont {J.}~\bibnamefont
  {Li}}, \bibinfo {author} {\bibfnamefont {Y.}~\bibnamefont {Shi}}, \ and\
  \bibinfo {author} {\bibfnamefont {X.}~\bibnamefont {Lu}},\ }\href {\doibase
  10.1103/PhysRevB.99.180504} {\bibfield  {journal} {\bibinfo  {journal} {Phys.
  Rev. B}\ }\textbf {\bibinfo {volume} {99}},\ \bibinfo {pages} {180504}
  (\bibinfo {year} {2019})}\BibitemShut {NoStop}%
\bibitem [{\citenamefont {Leng}\ \emph
  {et~al.}(2019{\natexlab{a}})\citenamefont {Leng}, \citenamefont {Orain},
  \citenamefont {Amato}, \citenamefont {Huang},\ and\ \citenamefont
  {de~Visser}}]{Leng2019}%
  \BibitemOpen
  \bibfield  {author} {\bibinfo {author} {\bibfnamefont {H.}~\bibnamefont
  {Leng}}, \bibinfo {author} {\bibfnamefont {J.-C.}\ \bibnamefont {Orain}},
  \bibinfo {author} {\bibfnamefont {A.}~\bibnamefont {Amato}}, \bibinfo
  {author} {\bibfnamefont {Y.~K.}\ \bibnamefont {Huang}}, \ and\ \bibinfo
  {author} {\bibfnamefont {A.}~\bibnamefont {de~Visser}},\ }\href {\doibase
  10.1103/PhysRevB.100.224501} {\bibfield  {journal} {\bibinfo  {journal}
  {Phys. Rev. B}\ }\textbf {\bibinfo {volume} {100}},\ \bibinfo {pages}
  {224501} (\bibinfo {year} {2019}{\natexlab{a}})}\BibitemShut {NoStop}%
\bibitem [{\citenamefont {Garcia-Campos}\ \emph {et~al.}(2021)\citenamefont
  {Garcia-Campos}, \citenamefont {Huang}, \citenamefont {de~Visser},\ and\
  \citenamefont {Hasselbach}}]{Garcia-Campos2021}%
  \BibitemOpen
  \bibfield  {author} {\bibinfo {author} {\bibfnamefont {P.}~\bibnamefont
  {Garcia-Campos}}, \bibinfo {author} {\bibfnamefont {Y.~K.}\ \bibnamefont
  {Huang}}, \bibinfo {author} {\bibfnamefont {A.}~\bibnamefont {de~Visser}}, \
  and\ \bibinfo {author} {\bibfnamefont {K.}~\bibnamefont {Hasselbach}},\
  }\href {\doibase 10.1103/PhysRevB.103.104510} {\bibfield  {journal} {\bibinfo
   {journal} {Phys. Rev. B}\ }\textbf {\bibinfo {volume} {103}},\ \bibinfo
  {pages} {104510} (\bibinfo {year} {2021})}\BibitemShut {NoStop}%
\bibitem [{\citenamefont {Furue}\ \emph {et~al.}(2021)\citenamefont {Furue},
  \citenamefont {Fujino}, \citenamefont {Salis}, \citenamefont {Leng},
  \citenamefont {Ishikawa}, \citenamefont {Naka}, \citenamefont {Nakano},
  \citenamefont {Huang}, \citenamefont {de~Visser},\ and\ \citenamefont
  {Ohmura}}]{Furue2021}%
  \BibitemOpen
  \bibfield  {author} {\bibinfo {author} {\bibfnamefont {Y.}~\bibnamefont
  {Furue}}, \bibinfo {author} {\bibfnamefont {T.}~\bibnamefont {Fujino}},
  \bibinfo {author} {\bibfnamefont {M.~V.}\ \bibnamefont {Salis}}, \bibinfo
  {author} {\bibfnamefont {H.}~\bibnamefont {Leng}}, \bibinfo {author}
  {\bibfnamefont {F.}~\bibnamefont {Ishikawa}}, \bibinfo {author}
  {\bibfnamefont {T.}~\bibnamefont {Naka}}, \bibinfo {author} {\bibfnamefont
  {S.}~\bibnamefont {Nakano}}, \bibinfo {author} {\bibfnamefont
  {Y.}~\bibnamefont {Huang}}, \bibinfo {author} {\bibfnamefont
  {A.}~\bibnamefont {de~Visser}}, \ and\ \bibinfo {author} {\bibfnamefont
  {A.}~\bibnamefont {Ohmura}},\ }\href {\doibase 10.1103/PhysRevB.104.144510}
  {\bibfield  {journal} {\bibinfo  {journal} {Phys. Rev. B}\ }\textbf {\bibinfo
  {volume} {104}},\ \bibinfo {pages} {144510} (\bibinfo {year}
  {2021})}\BibitemShut {NoStop}%
\bibitem [{\citenamefont {Leng}\ \emph
  {et~al.}(2019{\natexlab{b}})\citenamefont {Leng}, \citenamefont {Ohmura},
  \citenamefont {Anh}, \citenamefont {Ishikawa}, \citenamefont {Naka},
  \citenamefont {Huang},\ and\ \citenamefont {de~Visser}}]{leng2019p}%
  \BibitemOpen
  \bibfield  {author} {\bibinfo {author} {\bibfnamefont {H.}~\bibnamefont
  {Leng}}, \bibinfo {author} {\bibfnamefont {A.}~\bibnamefont {Ohmura}},
  \bibinfo {author} {\bibfnamefont {L.~N.}\ \bibnamefont {Anh}}, \bibinfo
  {author} {\bibfnamefont {F.}~\bibnamefont {Ishikawa}}, \bibinfo {author}
  {\bibfnamefont {T.}~\bibnamefont {Naka}}, \bibinfo {author} {\bibfnamefont
  {Y.~K.}\ \bibnamefont {Huang}}, \ and\ \bibinfo {author} {\bibfnamefont
  {A.}~\bibnamefont {de~Visser}},\ }\href {\doibase 10.1088/1361-648x/ab49b5}
  {\bibfield  {journal} {\bibinfo  {journal} {Journal of Physics: Condensed
  Matter}\ }\textbf {\bibinfo {volume} {32}},\ \bibinfo {pages} {025603}
  (\bibinfo {year} {2019}{\natexlab{b}})}\BibitemShut {NoStop}%
\bibitem [{\citenamefont {Kudo}\ \emph {et~al.}(2016)\citenamefont {Kudo},
  \citenamefont {Ishii},\ and\ \citenamefont {Nohara}}]{Kudo2016}%
  \BibitemOpen
  \bibfield  {author} {\bibinfo {author} {\bibfnamefont {K.}~\bibnamefont
  {Kudo}}, \bibinfo {author} {\bibfnamefont {H.}~\bibnamefont {Ishii}}, \ and\
  \bibinfo {author} {\bibfnamefont {M.}~\bibnamefont {Nohara}},\ }\href
  {\doibase 10.1103/PhysRevB.93.140505} {\bibfield  {journal} {\bibinfo
  {journal} {Phys. Rev. B}\ }\textbf {\bibinfo {volume} {93}},\ \bibinfo
  {pages} {140505} (\bibinfo {year} {2016})}\BibitemShut {NoStop}%
\bibitem [{\citenamefont {Ryu}(2015)}]{Ryu2015}%
  \BibitemOpen
  \bibfield  {author} {\bibinfo {author} {\bibfnamefont {G.}~\bibnamefont
  {Ryu}},\ }\href {https://link.springer.com/article/10.1007/s10948-015-3195-2}
  {\bibfield  {journal} {\bibinfo  {journal} {Journal of Superconductivity and
  Novel Magnetism}\ }\textbf {\bibinfo {volume} {28}},\ \bibinfo {pages} {3275}
  (\bibinfo {year} {2015})}\BibitemShut {NoStop}%
\bibitem [{\citenamefont {Hooda}\ and\ \citenamefont
  {Yadav}(2018)}]{Hooda2018}%
  \BibitemOpen
  \bibfield  {author} {\bibinfo {author} {\bibfnamefont {M.~K.}\ \bibnamefont
  {Hooda}}\ and\ \bibinfo {author} {\bibfnamefont {C.~S.}\ \bibnamefont
  {Yadav}},\ }\href {\doibase 10.1209/0295-5075/121/17001} {\bibfield
  {journal} {\bibinfo  {journal} {Europhysics Letters}\ }\textbf {\bibinfo
  {volume} {121}},\ \bibinfo {pages} {17001} (\bibinfo {year}
  {2018})}\BibitemShut {NoStop}%
\bibitem [{\citenamefont {Vasdev}\ \emph {et~al.}(2019)\citenamefont {Vasdev},
  \citenamefont {Sirohi}, \citenamefont {Hooda}, \citenamefont {Yadav},\ and\
  \citenamefont {Sheet}}]{Vasdev2019}%
  \BibitemOpen
  \bibfield  {author} {\bibinfo {author} {\bibfnamefont {A.}~\bibnamefont
  {Vasdev}}, \bibinfo {author} {\bibfnamefont {A.}~\bibnamefont {Sirohi}},
  \bibinfo {author} {\bibfnamefont {M.~K.}\ \bibnamefont {Hooda}}, \bibinfo
  {author} {\bibfnamefont {C.~S.}\ \bibnamefont {Yadav}}, \ and\ \bibinfo
  {author} {\bibfnamefont {G.}~\bibnamefont {Sheet}},\ }\href {\doibase
  10.1088/1361-648x/ab49b5} {\bibfield  {journal} {\bibinfo  {journal} {Journal
  of Physics: Condensed Matter}\ }\textbf {\bibinfo {volume} {32}},\ \bibinfo
  {pages} {125701} (\bibinfo {year} {2019})}\BibitemShut {NoStop}%
\bibitem [{\citenamefont {Lyons}\ \emph {et~al.}(1976)\citenamefont {Lyons},
  \citenamefont {Schleich},\ and\ \citenamefont {Wold}}]{Lyons1976}%
  \BibitemOpen
  \bibfield  {author} {\bibinfo {author} {\bibfnamefont {A.}~\bibnamefont
  {Lyons}}, \bibinfo {author} {\bibfnamefont {D.}~\bibnamefont {Schleich}}, \
  and\ \bibinfo {author} {\bibfnamefont {A.}~\bibnamefont {Wold}},\ }\href@noop
  {} {\bibfield  {journal} {\bibinfo  {journal} {Mat. Res. Bull.}\ }\textbf
  {\bibinfo {volume} {11}},\ \bibinfo {pages} {1155} (\bibinfo {year}
  {1976})}\BibitemShut {NoStop}%
\bibitem [{Sup()}]{Supp}%
  \BibitemOpen
  \href@noop {} {\bibinfo  {journal} {See Supplemental Material https:...}\
  }\BibitemShut {NoStop}%
\bibitem [{\citenamefont {Stewart}(1983)}]{Stewart1983}%
  \BibitemOpen
\bibfield  {journal} {  }\bibfield  {author} {\bibinfo {author} {\bibfnamefont
  {G.~R.}\ \bibnamefont {Stewart}},\ }\href {\doibase 10.1063/1.1137207}
  {\bibfield  {journal} {\bibinfo  {journal} {Review of Scientific
  Instruments}\ }\textbf {\bibinfo {volume} {54}},\ \bibinfo {pages} {1}
  (\bibinfo {year} {1983})}\BibitemShut {NoStop}%
\bibitem [{\citenamefont {Hein}\ and\ \citenamefont {Falge}(1961)}]{Hein1961}%
  \BibitemOpen
  \bibfield  {author} {\bibinfo {author} {\bibfnamefont {R.~A.}\ \bibnamefont
  {Hein}}\ and\ \bibinfo {author} {\bibfnamefont {R.~L.}\ \bibnamefont
  {Falge}},\ }\href {\doibase 10.1103/PhysRev.123.407} {\bibfield  {journal}
  {\bibinfo  {journal} {Phys. Rev.}\ }\textbf {\bibinfo {volume} {123}},\
  \bibinfo {pages} {407} (\bibinfo {year} {1961})}\BibitemShut {NoStop}%
\bibitem [{\citenamefont {Werthamer}\ \emph {et~al.}(1966)\citenamefont
  {Werthamer}, \citenamefont {Helfand},\ and\ \citenamefont
  {Hohenberg}}]{Werthamer1966}%
  \BibitemOpen
  \bibfield  {author} {\bibinfo {author} {\bibfnamefont {N.~R.}\ \bibnamefont
  {Werthamer}}, \bibinfo {author} {\bibfnamefont {E.}~\bibnamefont {Helfand}},
  \ and\ \bibinfo {author} {\bibfnamefont {P.~C.}\ \bibnamefont {Hohenberg}},\
  }\href@noop {} {\bibfield  {journal} {\bibinfo  {journal} {Phys. Rev.}\
  }\textbf {\bibinfo {volume} {147}},\ \bibinfo {pages} {295} (\bibinfo {year}
  {1966})}\BibitemShut {NoStop}%
\bibitem [{\citenamefont {Timmons}\ \emph {et~al.}(2020)\citenamefont
  {Timmons}, \citenamefont {Teknowijoyo}, \citenamefont
  {Ko\ifmmode~\acute{n}\else \'{n}\fi{}czykowski}, \citenamefont {Cavani},
  \citenamefont {Tanatar}, \citenamefont {Ghimire}, \citenamefont {Cho},
  \citenamefont {Lee}, \citenamefont {Ke}, \citenamefont {Jo}, \citenamefont
  {Bud'ko}, \citenamefont {Canfield}, \citenamefont {Orth}, \citenamefont
  {Scheurer},\ and\ \citenamefont {Prozorov}}]{Timmons2020}%
  \BibitemOpen
  \bibfield  {author} {\bibinfo {author} {\bibfnamefont {E.~I.}\ \bibnamefont
  {Timmons}}, \bibinfo {author} {\bibfnamefont {S.}~\bibnamefont
  {Teknowijoyo}}, \bibinfo {author} {\bibfnamefont {M.}~\bibnamefont
  {Ko\ifmmode~\acute{n}\else \'{n}\fi{}czykowski}}, \bibinfo {author}
  {\bibfnamefont {O.}~\bibnamefont {Cavani}}, \bibinfo {author} {\bibfnamefont
  {M.~A.}\ \bibnamefont {Tanatar}}, \bibinfo {author} {\bibfnamefont
  {S.}~\bibnamefont {Ghimire}}, \bibinfo {author} {\bibfnamefont
  {K.}~\bibnamefont {Cho}}, \bibinfo {author} {\bibfnamefont {Y.}~\bibnamefont
  {Lee}}, \bibinfo {author} {\bibfnamefont {L.}~\bibnamefont {Ke}}, \bibinfo
  {author} {\bibfnamefont {N.~H.}\ \bibnamefont {Jo}}, \bibinfo {author}
  {\bibfnamefont {S.~L.}\ \bibnamefont {Bud'ko}}, \bibinfo {author}
  {\bibfnamefont {P.~C.}\ \bibnamefont {Canfield}}, \bibinfo {author}
  {\bibfnamefont {P.~P.}\ \bibnamefont {Orth}}, \bibinfo {author}
  {\bibfnamefont {M.~S.}\ \bibnamefont {Scheurer}}, \ and\ \bibinfo {author}
  {\bibfnamefont {R.}~\bibnamefont {Prozorov}},\ }\href {\doibase
  10.1103/PhysRevResearch.2.023140} {\bibfield  {journal} {\bibinfo  {journal}
  {Phys. Rev. Research}\ }\textbf {\bibinfo {volume} {2}},\ \bibinfo {pages}
  {023140} (\bibinfo {year} {2020})}\BibitemShut {NoStop}%
\bibitem [{\citenamefont {Tinkham}(1996)}]{Tinkham1996}%
  \BibitemOpen
  \bibfield  {author} {\bibinfo {author} {\bibfnamefont {M.}~\bibnamefont
  {Tinkham}},\ }\href@noop {} {\emph {\bibinfo {title} {Introduction to
  Superconductivity}}}\ (\bibinfo  {publisher} {McGraw-Hill Inc., New York},\
  \bibinfo {year} {1996})\BibitemShut {NoStop}%
\bibitem [{\citenamefont {Poole}\ \emph {et~al.}(2007)\citenamefont {Poole},
  \citenamefont {Farach}, \citenamefont {Creswick},\ and\ \citenamefont
  {Prozorov}}]{Poole2007}%
  \BibitemOpen
  \bibfield  {author} {\bibinfo {author} {\bibfnamefont {C.~P.}\ \bibnamefont
  {Poole}}, \bibinfo {author} {\bibfnamefont {H.~A.}\ \bibnamefont {Farach}},
  \bibinfo {author} {\bibfnamefont {R.~J.}\ \bibnamefont {Creswick}}, \ and\
  \bibinfo {author} {\bibfnamefont {R.}~\bibnamefont {Prozorov}},\ }\href@noop
  {} {\emph {\bibinfo {title} {Superconductivity}}},\ \bibinfo {edition} {2nd}\
  ed.\ (\bibinfo  {publisher} {Elsevier},\ \bibinfo {address} {Amsterdam},\
  \bibinfo {year} {2007})\BibitemShut {NoStop}%
\bibitem [{\citenamefont {Samoilenka}\ and\ \citenamefont
  {Babaev}(2021)}]{Samoilenka2021}%
  \BibitemOpen
  \bibfield  {author} {\bibinfo {author} {\bibfnamefont {A.}~\bibnamefont
  {Samoilenka}}\ and\ \bibinfo {author} {\bibfnamefont {E.}~\bibnamefont
  {Babaev}},\ }\href {\doibase 10.1103/PhysRevB.103.224516} {\bibfield
  {journal} {\bibinfo  {journal} {Phys. Rev. B}\ }\textbf {\bibinfo {volume}
  {103}},\ \bibinfo {pages} {224516} (\bibinfo {year} {2021})}\BibitemShut
  {NoStop}%
\end{thebibliography}
\end{document}